\documentclass[]{aastex63}
\hypersetup{linkcolor=blue, urlcolor=blue}
\usepackage{graphicx}
\usepackage{amsmath}
\usepackage{lineno}
\usepackage{subfigure}
\usepackage{graphicx}
\usepackage{hyperref}
\usepackage{appendix}
\usepackage{cleveref}
\usepackage{lineno}
\begin{document}
\title{Magnetically Aligned Striations in the L914 Filamentary Cloud}

%============================================================
%======== Abstract

\begin{abstract}
We present CO\,($J$ = 1--0) multi-line observations toward the L914 dark cloud in the vicinity of the Cygnus X region, 
using the 13.7\,m millimeter telescope of the Purple Mountain Observatory (PMO). The CO observations reveal in the 
L914 cloud a long filament with an angular length of $\sim$\,3$\fdg$6, corresponding to approximately 50\,pc at the 
measured distance of $\sim$\,760\,pc. 
Furthermore, a group of hair-like striations are discovered in the two subregions of the L914 cloud, which are connected 
with the dense ridge of the filament. These striations display quasi-periodic characteristics in both the CO intensity images and position-velocity diagrams.
Two of the striations also show increasing velocity gradients and dispersions toward the 
dense ridge, which could be fitted by accretion flows under gravity.
Based on the $Planck$ 353 GHz dust polarization data, we find that the striations are well aligned with the magnetic fields. 
Moreover, both the striations and magnetic fields are perpendicular to the dense ridge, which constructs a bimodal configuration. 
Using the classic method, we estimate the strength of magnetic field, and further evaluate the relative importance of gravity, 
turbulence and magnetic field, and find that the L914 cloud is strongly magnetized. 
Our results suggest that magnetic fields play an important role in the formation of filamentary structures by channelling the 
material along the striations toward the dense ridge. 
The comparison between the observations and simulations suggests that striations could be a product of the magnetohydrodynamic 
(MHD) process.

\end{abstract}


%============================================================
%======== Authors

\author{Li Sun}
\affil{Purple Mountain Observatory, and Key Laboratory of Radio Astronomy, Chinese Academy of Sciences, \\10 Yuanhua Road, Nanjing 210023, China; xpchen@pmo.ac.cn, lisun@pmo.ac.cn}
\affil{School of Astronomy and Space Science, University of Science and Technology of China, Hefei, Anhui 230023, China}

\author{Xuepeng Chen}
\affil{Purple Mountain Observatory, and Key Laboratory of Radio Astronomy, Chinese Academy of Sciences, \\10 Yuanhua Road, Nanjing 210023, China; xpchen@pmo.ac.cn, lisun@pmo.ac.cn}
\affil{School of Astronomy and Space Science, University of Science and Technology of China, Hefei, Anhui 230023, China}

\author{Min Fang}
\affil{Purple Mountain Observatory, and Key Laboratory of Radio Astronomy, Chinese Academy of Sciences, \\10 Yuanhua Road, Nanjing 210023, China; xpchen@pmo.ac.cn, lisun@pmo.ac.cn}
\affil{School of Astronomy and Space Science, University of Science and Technology of China, Hefei, Anhui 230023, China}

\author{Shaobo Zhang}
\affil{Purple Mountain Observatory, and Key Laboratory of Radio Astronomy, Chinese Academy of Sciences, \\10 Yuanhua Road, Nanjing 210023, China; xpchen@pmo.ac.cn, lisun@pmo.ac.cn}

\author{Yan Gong}
\affil{Purple Mountain Observatory, and Key Laboratory of Radio Astronomy, Chinese Academy of Sciences, \\10 Yuanhua Road, Nanjing 210023, China; xpchen@pmo.ac.cn, lisun@pmo.ac.cn}

\author{Jiancheng Feng}
\affil{Purple Mountain Observatory, and Key Laboratory of Radio Astronomy, Chinese Academy of Sciences, \\10 Yuanhua Road, Nanjing 210023, China; xpchen@pmo.ac.cn, lisun@pmo.ac.cn}
\affil{School of Astronomy and Space Science, University of Science and Technology of China, Hefei, Anhui 230023, China}

\author{Xuefu Li}
\affil{School of Physics and Astronomy, Sun Yat-Sen University, Zhuhai campus, No. 2, Daxue Road, Zhuhai, Guangdong, 519082, China}

\author{Qing-Zeng Yan}
\affil{Purple Mountain Observatory, and Key Laboratory of Radio Astronomy, Chinese Academy of Sciences, \\10 Yuanhua Road, Nanjing 210023, China; xpchen@pmo.ac.cn, lisun@pmo.ac.cn}

\author{Ji Yang}
\affil{Purple Mountain Observatory, and Key Laboratory of Radio Astronomy, Chinese Academy of Sciences, \\10 Yuanhua Road, Nanjing 210023, China; xpchen@pmo.ac.cn, lisun@pmo.ac.cn}
\affil{School of Astronomy and Space Science, University of Science and Technology of China, Hefei, Anhui 230023, China}

%============================================================
%======== Introduction

\section{introduction} \label{sec:intro}
The universality of filamentary structures in molecular clouds has been revealed by multi-wavelength surveys (see, e.g., \citealt{Schneider1979, Molinari2010, Schuller2017, Yuan2021}). $Herschel$ observations suggest that filaments play an important role in connecting molecular clouds to star formation (\citealt{Andre2014}). Nevertheless, the physical mechanisms relevant to the formation and evolution of filaments are complex. The interaction of gravity, turbulence, thermal pressure and magnetic field and their relative importance remain a subject of debate (see, e.g., \citealt{Inutsuka1992, Padoan2001, Hennebelle2013, Gomez2014}).

The large-scale ordered striations may help us to understand the physical processes leading to the formation of filamentary structures (see reviews in \citealt{Hacar2023} and \citealt{Pineda2023}). Striations were first identified by \citet{Goldsmith2008}, who discovered a string of hair-like structures with a large dynamic range CO mapping in the Taurus region. The polarization measurements reveal that these striations are well aligned with the projected magnetic field. The subsequent $Herschel$ observations at the same region found the perpendicular striations connecting to the crest of the B211 filament and led to the view that the filament is accreting material through the striations \citep{Palmeirim2013, Shimajiri2019}. The molecular striations were also discovered in the Musca \citep{Cox2016}, L1642 \citep{Malinen2014, Malinen2016} and Polaris \citep{Panopoulou2016} clouds, etc.

In addition to molecular clouds, striations are also observed in the diffuse atomic medium (e.g., \citealt{McClure2006, Tritsis2019}). These structures are observed to be ordered, quasi-periodic and well aligned with the plane-of-sky (POS) magnetic field (e.g., \citealt{Heyer2008, Heyer2016, Panopoulou2016, Soler2019}). With these striking features, it is natural to conjecture that the striation structures are the result of material flowing along magnetic field lines \citep{Goldsmith2008, Palmeirim2013}. The analysis of velocity fields and line excitation in Taurus suggests either Kelvin-Helmholtz instability or magnetohydrodynamic (MHD) wave as the origin of Taurus striations \citep{Heyer2016}. Alternatively, \cite{Chen2017} proposed that the striations are the corrugations of sheets caused by the thin shell instability. Nevertheless, due to the extremely small sample of the observed striations, it is still difficult to set strong constraints on the theoretical models and simulations. Therefore, it is of importance to search for more filaments associated with striations, in order to better understand the formation mechanism of filaments.

The L914 cloud is a dark nebula first listed in \citet{Lynds1962}. It was covered by the $\rm ^{13}CO$ survey with the two $\rm 4~m$ millimeter telescopes at Nagoya University \citep{Dobashi1994}. In the $^{13}$CO observations, the L914 cloud showed an elongated structure overall. In this work, we present higher-resolution CO observations toward the L914 cloud, as part of the Milky Way Imaging Scroll Painting (MWISP) project\footnote{\url{http://www.radioast.nsdc.cn/yhhjindex.php}}, which is an unbiased CO $(J = 1-0)$ multi-line survey toward the northern Galactic plane using the 13.7\,m millimeter telescope of the Purple Mountain Observatory (PMO; \citealt{Su2019, Sun2021} ). Additionally, complementary dust polarization data from the $Planck$ are also used. The observations and data reduction are introduced in Section 2. In Section 3, we present observational results and report the discovery of a long filament associated with magnetically aligned striations. In Section 4, we evaluate the balance between the magnetic field, turbulence and gravity in the L914 cloud and discuss the potential mechanisms involved in the formation of the filamentary structures, as well as diffuse striations. The main conclusions are summarized in Section 5.

%============================================================
%======== Observavtion

\section{Observations}

\subsection{PMO 13.7\,m CO observations} 
The CO observations toward the L914 cloud covered a region within $81\arcdeg \le l \le 84\arcdeg$ and $-3\arcdeg \le b \le -1\arcdeg$. The observations were conducted with the PMO 13.7\,m millimeter telescope located in Delingha, China, from March 2012 to May 2018. The nine-beam Superconducting Spectroscopic Array Receiver (SSAR; \citealt{Shan2012}) was used as the front end, and the $\rm ^{12}CO$, $\rm^{13}CO$, and $\rm C^{18}O$ $J = 1-0$ emission lines were simultaneously observed in the sideband separation mode. The upper sideband (USB) contains the $\rm^{12}CO$ line and the lower sideband (LSB) contains the $\rm^{13}CO$ and $\rm C^{18}O$ lines. The half-power beamwidth (HPBW) was $\sim55\arcsec$ at $\rm 110~GHz$ and $\sim52\arcsec$ at $\rm 115~GHz$. The typical system temperature was $\rm\sim 275~K$ for $\rm ^{12}CO$ and $\rm\sim 155~K$ for $\rm ^{13}CO$ and $\rm C^{18}O$. A Fast Fourier Transform Spectrometer (FFTS) was used as the back end, which has a total bandwidth of $\rm 1~GHz$ and contains 16384 channels. The corresponding velocity resolution is $\rm 0.16~km~s^{-1}$ for the $\rm ^{12}CO$ line and $\rm 0.17~km~s^{-1}$ for the $\rm ^{13}CO$ and $\rm C^{18}O$ lines, respectively. The observed area is divided into individual $\rm 30\arcmin \times 30\arcmin$ cells. Each cell was observed in the OTF mode with a scanning rate of $50\arcsec$ per second and a dump time of $\rm 0.3~s$. The scanning interval was $15\arcsec$ ($\rm 50\arcsec~s^{-1} \times 0.3~s$). To reduce the scanning effects, each cell was mapped along both the Galactic longitude and latitude.

We calibrated the antenna temperature ($T_{\rm A}$) with the standard chopper-wheel method \citep{Ulich1976}. The main-beam temperature ($T_{\rm mb}$) was derived from the antenna temperature ($T_{\rm A}$) using the equation of $T_{\rm mb} = T_{\rm A}/B_{\rm eff}$, where the main-beam efficiencies ($B_{\rm eff}$) were approximately 44\% for the USB and 48\% for the LSB during the observations. The calibration errors were estimated to be within 10\%.

The raw data were reduced by the MWISP working group with the GILDAS software\footnote{\url{http://iram.fr/IRAMFR/GILDAS}} and self-developed pipelines. After eliminating the abnormal data, we mosaicked the data cubes ($30\arcmin \times 30\arcmin$ FITS) of cloud regions and regridded the maps into $30\arcsec \times 30\arcsec$. The typical rms noise level was below $\rm \sim0.5~K$ for $\rm ^{12}CO$ at a channel width of $\rm 0.16~km~s^{-1}$, and $\rm \sim0.3~K$ for $\rm ^{13}CO$ and $\rm C^{18}O$ at a channel width of $\rm 0.17~km~s^{-1}$. All velocities given in this work are relative to the local standard of rest (LSR).

\subsection{$Planck$ $353~GHz$ observation}\label{sec:o2}
To investigate the magnetic field of the L914 cloud, we retrieved dust polarization data from the $Planck$ Legacy Archive\footnote{\url{http://pla.esac.esa.int}}. The $Planck$ all-sky survey \citep{Planck2011} observed the linear polarization in seven bands from 30 to $\rm 353~GHz$, of which the $\rm 353~GHz$ band was the most sensitive for detecting dust polarization \citep{Planck2015}. Additionally, the cosmic microwave background (CMB) does not significantly contribute to polarized emission in this band when observing molecular clouds \citep{Soler2019}. Therefore, $Planck$ $\rm 353~GHz$ sub-millimeter polarization data were used in this work to trace magnetic fields. The original maps have a resolution of $4\farcm8$ in the HEALPIX\footnote{\url{http://healpix.sf.net}} format with a pixelization at $N_{side}$ = 2048, which corresponds to the pixel size of $1\farcm7$. We smoothed the maps into $10\arcmin$ to increase the signal-to-noise ratio (S/N) of the extended regions, as suggested by \citet{Planck_2016}. The polarization position angle (PA) is calculated with
\begin{equation}
\theta_{\rm PA} = -\frac{1}{2}~arctan(\frac{U}{Q}),
\end{equation} 
where $\theta_{\rm PA}$ is given in the IAU convention, i.e., $\theta_{\rm PA}=0\arcdeg$ points to the north and $\theta_{\rm PA}$ increases anticlockwise. The orientation of the magnetic field is perpendicular to the position angle, $\theta_{\rm B} = \theta_{\rm PA} - \frac{\pi}{2}$.

%============================================================
%========= Results

\section{Results}

%========= Results 1
\subsection{L914: A Filamentary Molecular Cloud with Striations}\label{sec:r1}
\begin{figure}
\centering
\includegraphics[scale=0.45]{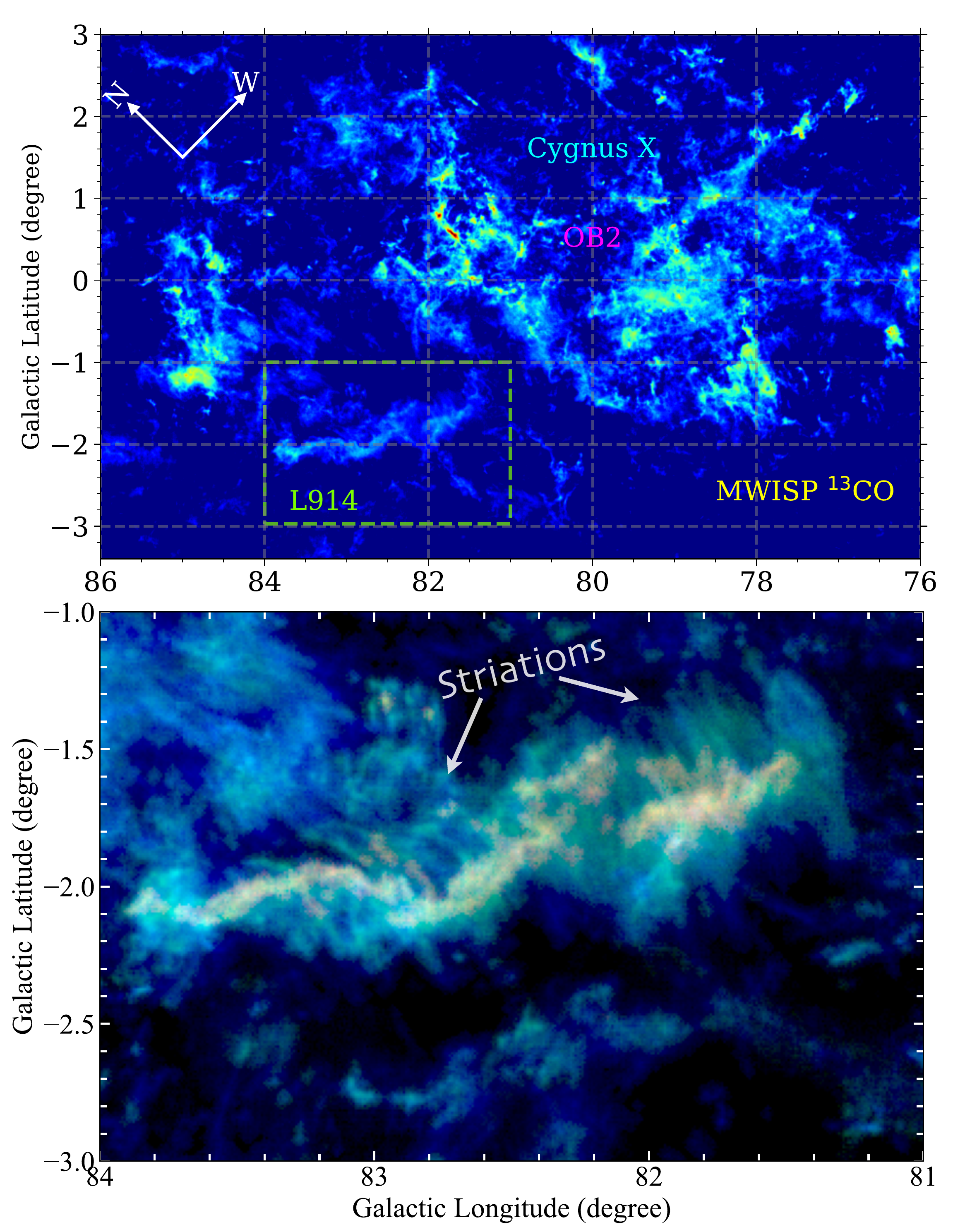}
\caption{Top: Overview of $\rm Cygnus~X$ region in the MWISP $\rm ^{13}CO$ observations \citep{zhang2023cygnus}, which is integrated over the velocity range of $\rm [-100,\, 40]~km~s^{-1}$. Bottom: Three-color image of the $\rm ^{12}CO$ (blue), $\rm ^{13}CO$ (green) and $\rm C^{18}O$ (red) emission toward the L914 cloud. The integral velocity intervals for the individual lines are the same as Figure~\ref{fig3}.}
\label{fig1}
\end{figure}

Figure~\ref{fig1} presents an overview of the $\rm Cygnus~X$ region in the MWISP $\rm ^{13}CO$ observations. The integral velocity interval is $\rm [-100, \, 40]~km~s^{-1}$. Around the center of the Cygnus X region (Cyg OB2 cluster), there are many bright filamentary molecular clouds, and a lot of research has focused on these molecular clouds over the past few decades (e.g., \citealt{Schneider2010, Cao2022, Lichong2023, Gong2023}). However, the relatively isolated L914 molecular cloud located at the periphery of this region has received little attention. The $\rm ^{13}CO$ $J=1-0$ survey from the Nagoya 4 m millimeter telescope ($\sim 2\farcm 7$ in angular resolution) showed an elongated structure of the L914 cloud at the velocity range of $\rm \sim [1, \, 6]~km~s^{-1}$ \citep{Dobashi1994}. The enlarged view in Figure~\ref{fig1} displays the three-color image of the MWISP CO data toward the L914 cloud, with $\rm ^{12}CO$ emission shown in blue, $\rm ^{13}CO$ in green, and $\rm C^{18}O$ in red. Based on the higher-angular resolution MWISP CO data, we reveal a long filamentary structure in the L914 cloud.

\begin{figure}
\centering
\includegraphics[scale=0.4]{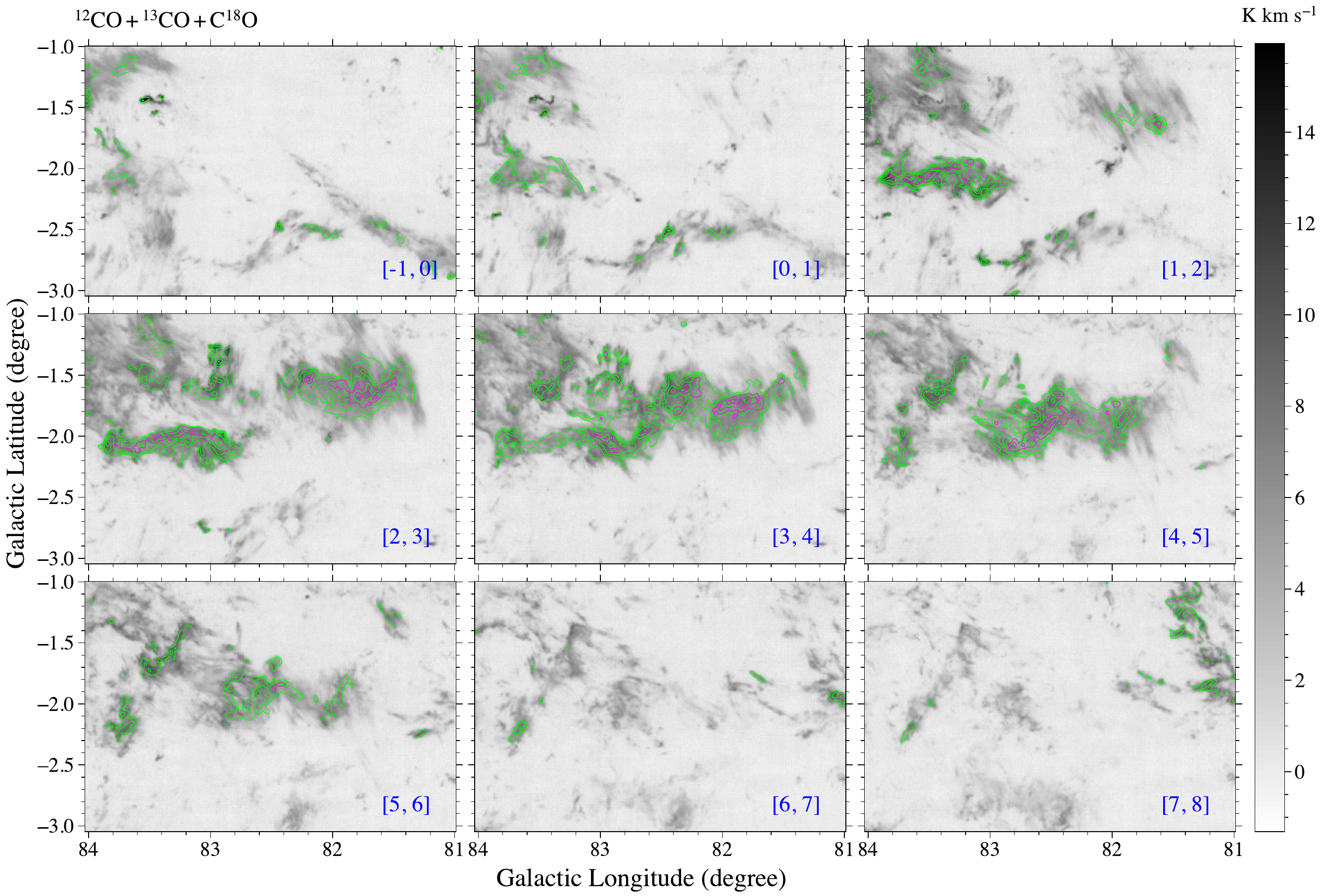}
\caption{Velocity-integrated channel maps of the $\rm ^{12}CO$ (gray-scale background), $\rm ^{13}CO$ (lime contours), and $\rm C^{18}O$ (magenta contours) line emission. The integrated velocity range is marked in the bottom right corner of each panel with the units of $\rm km~s^{-1}$. The $\rm ^{13}CO$ contours start at $\rm 1~K~km~s^{-1}$ and increase with a step of $\rm 1~K~km~s^{-1}$ ($\rm\sigma \sim 0.15~~K~km~s^{-1}$). The $\rm C^{18}O$ contours start at $\rm 0.5~K~km~s^{-1}$ and increase by $\rm 0.5~K~km~s^{-1}$ ($\rm \sigma \sim 0.15~~K~km~s^{-1}$).} 
\label{fig2}
\end{figure}

\begin{figure}
\centering
\includegraphics[scale=0.65]{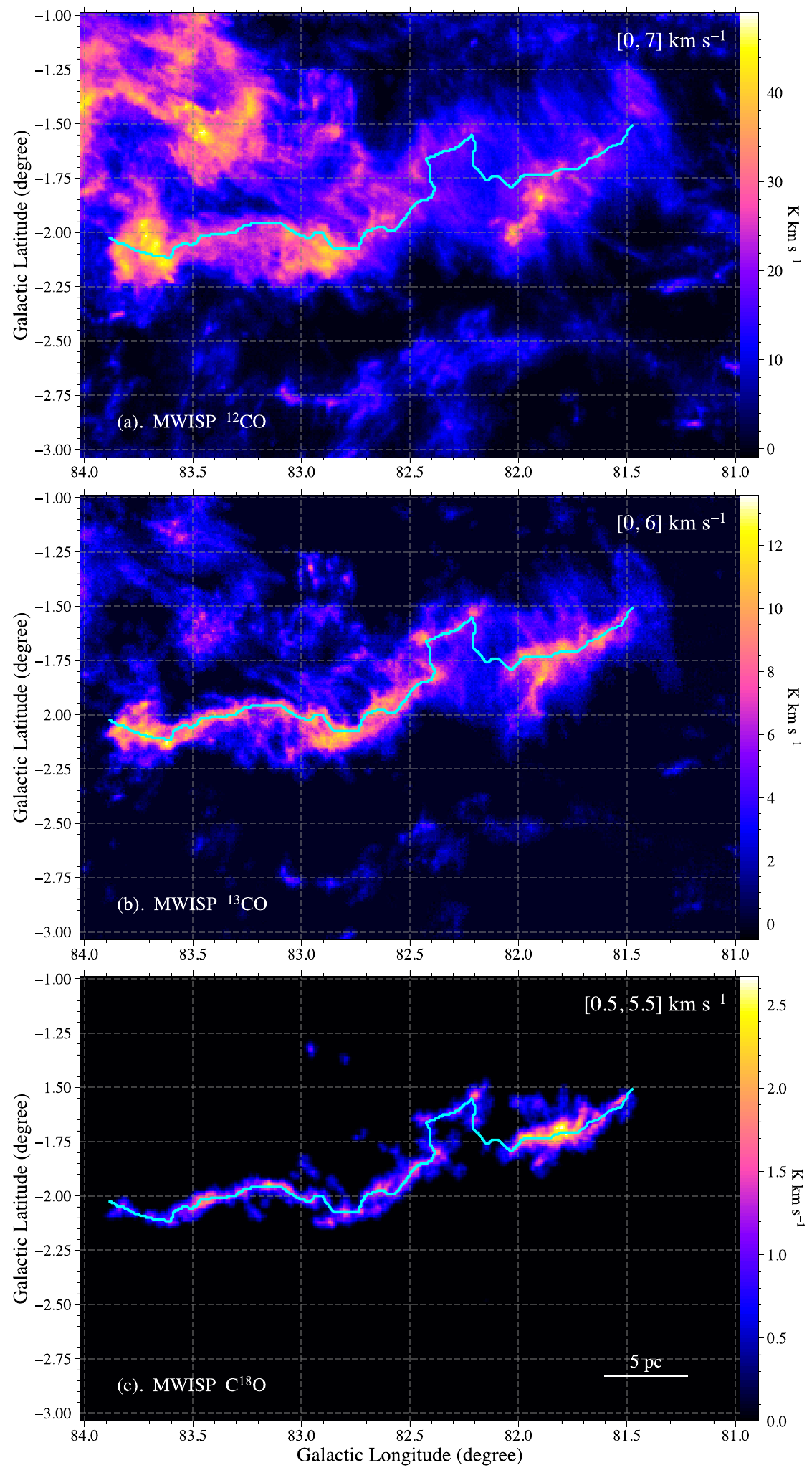}
\caption{Velocity-integrated intensity maps of $\rm ^{12}CO$ , $\rm ^{13}CO$ and $\rm C^{18}O$ (from top to bottom) toward the L914 cloud. The integral interval is labeled at the top right corner of each panel. The cyan line in each panel represents the ridge of the L914 filamentary cloud identified by the DisPerSE algorithm using the $\rm C^{18}O$ data.}
\label{fig3}
\end{figure}

\begin{figure}
\centering
\includegraphics[scale=0.64]{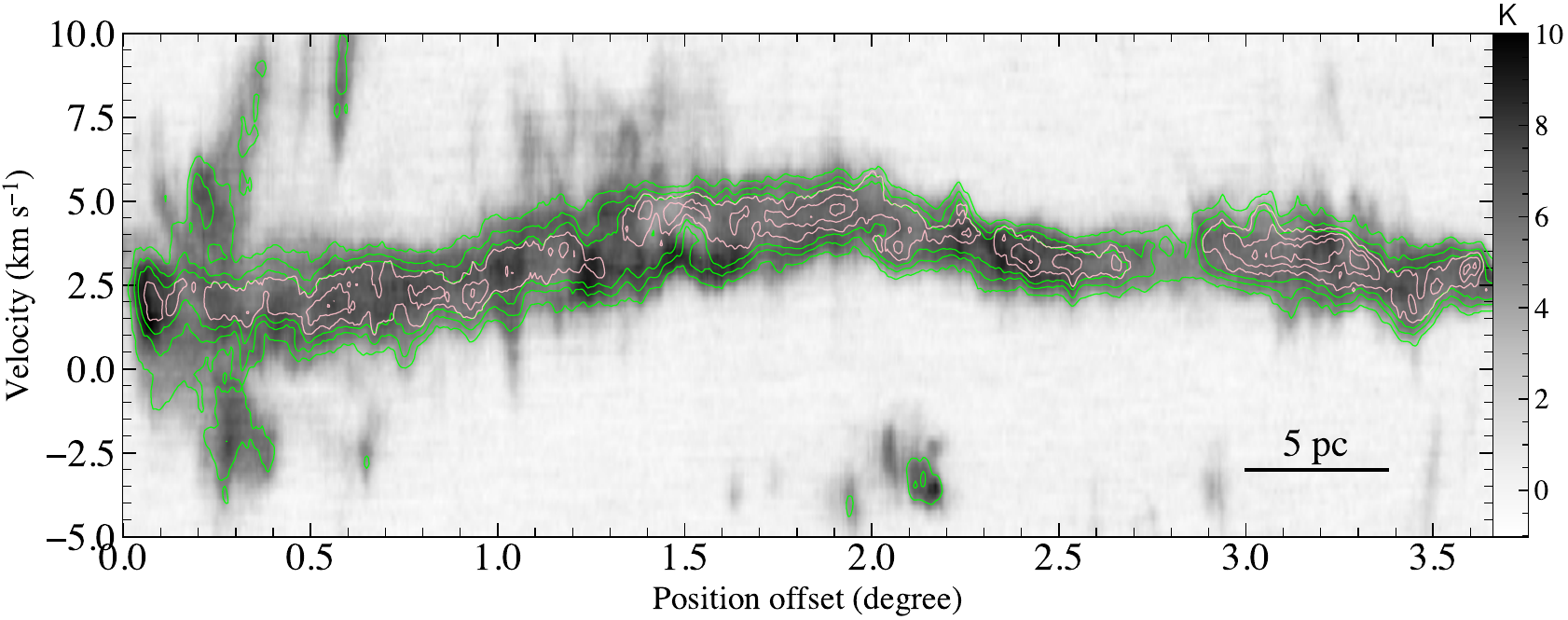}
\caption{CO position-velocity image along the ridge of the L914 filamentary cloud (from east to west). The gray-scale background represents the $\rm ^{12}CO$ emission. The lime and pink contours represent the $\rm ^{13}CO$ (levels = 0.75, 1.75, $\rm 2.75~K$; $\rm\sigma \sim 0.15~K$) and $\rm C^{18}O$ (levels = 0.5, 1.0, $\rm 1.5~K$; $\rm\sigma \sim 0.15~K$) emission, respectively.}
\label{fig4}
\end{figure}

Figure~\ref{fig2} shows the velocity channel maps of the MWISP CO emission toward the L914 cloud. The gray-scale background represents the $\rm ^{12}CO$ emission, with the overlapped green and magenta contours showing the $\rm ^{13}CO$ and $\rm C^{18}O$ emission, respectively. As shown in this image, the L914 cloud is detected in the velocity range from $\rm \sim0$ to $\rm 7~km~s^{-1}$. The two ends of the filament appear in the velocity channel ranging from roughly 1 to $\rm 4~km~s^{-1}$, whereas the central part is predominantly  observed in about $\rm[3,~6]~km~s^{-1}$. The integrated intensity maps of $\rm ^{12}CO$, $\rm ^{13}CO$, and $\rm C^{18}O$ are shown in Figure~\ref{fig3}, where the three molecular lines are integrated over the velocity ranges of [0, \, 7], [0,\,  6] and $\rm [0.5,\, 5.5]~km~s^{-1}$, respectively. Compared with $^{12}$CO and $^{13}$CO, the $\rm C^{18}O$ line traces much denser regions. Figure~\ref{fig3}c shows that the L914 cloud has strong $\rm C^{18}O$ emission along the filament. We adopt the Discrete Persistent Structure Extractor (DisPerSE; \citealt{Sousbie2011}) algorithm on the $\rm C^{18}O$ data for the dense ridge extraction. As the DisPerSE algorithm extracts persistent structures by connecting topological critical points (e.g., maximum, minimum and saddle points), it is susceptible to local extrema. We thus smooth the $\rm C^{18}O$ data to $75\arcsec$ before extraction. In DisPerSE, the persistence and robustness thresholds are set to be 2 and $\rm 0.8~K~km~s^{-1}$ ($\rm \sigma\sim0.25~K~km~s^{-1}$,  see DisPerSE\footnote{\url{http://www2.iap.fr/users/sousbie/web/html/index55a0.html?category/Quick-start}} website for more details). The results obtained from DisPerSE are further visually inspected in the integrated intensity map and minor adjustments have been applied according to the $3\sigma$ contour. The resulting ridge is depicted in Figure~\ref{fig3} with solid cyan lines. Figure~\ref{fig4} presents the position-velocity (PV) map along the ridge from left to right. The structure extends more than $3\fdg6$ in length (i.e., $\rm \sim 50~pc$ at a distance of $\rm 760~pc$; see below), showing continuous and strong emission in both the $\rm ^{12}CO$ and $\rm ^{13}CO$ lines.

Interestingly, as seen in Figure~\ref{fig2}, distinct $\rm ^{12}CO$ and $\rm ^{13}CO$ hair-like structures are discovered in the velocity channels of [4, 5] and $\rm [1, \,2]~km~s^{-1}$, located in the central and right parts of the L914 cloud, respectively. We zoom in these two parts, and carefully choose the velocity ranges that could clearly detect the structures for integration. The $\rm ^{12}CO$ narrow-velocity integrated intensity maps are shown in Figure~\ref{fig5}, with the integrated velocity ranges of [4,\,6]  and $\rm [0,\, 2] ~km~s^{-1}$, respectively. As seen below, these hair-like structures display quasi-periodic characteristics and also parallel to the magnetic lines derived from the $Planck$ dust polarization (see Section~\ref{sec:r31}), and therefore identified as striations in this work (see definition suggested by \citealt{Hacar2023}). 

To investigate the periodicity of the striations, we present the profiles of the averaged integrated intensity along the directions (the arrow lines in Figure~\ref{fig5}) perpendicular to the striations in Figure~\ref{fig6}. The blue solid lines are the original averaged intensity and the dotted orange lines show the profiles after smoothing. We mark the local peaks of the smoothed intensity as the positions of the $\rm ^{12}CO$ striations. In the central part, four $\rm ^{12}CO$ striations are observed and it is evident that the spacing between the striations gradually decreases from left to right, ranging from about 1.4 to 0.6 pc, with a mean value of $\rm\sim 1.1~pc$. In the right part, six $\rm ^{12}CO$ striations are found to show a quasi-periodic pattern with a period T of about $\rm 1.4~pc$. Given that the striations are likely in a plane inclined to the plane of the sky by an angle of $\theta$, the actual period is determined as $\rm T/cos(\theta$). Here, we only set a lower limit for the oscillation period. Additionally, in the right part, quasi-periodic oscillation is also observed in the velocity dispersions of both the $\rm ^{12}CO$ and $\rm ^{13}CO$ spectra (see Figure~\ref{fig7}). The red dashed lines in Figure~\ref{fig7} show the sinusoidal function with a period of $\rm \sim 1.4~pc$. We assign the $\rm ^{12}CO$ striations S1-S10 from left to right. The contrast between the striation and its surroundings can be estimated with $\frac{I_{\rm max} - I_{\rm min}}{I_{\rm max}}$, where $I_{\rm max}$ and  $I_{\rm min}$ represent the local maximum and minimum of integral intensity, respectively. The mean contrasts in these two parts are $\sim 12\%$ and 28\%, respectively. Since the $\rm ^{12}CO$ emission is relatively diffuse, it is difficult to obtain the path of the striation skeleton. We thus choose the striations with prominent $\rm ^{13}CO$ emission to do the ridge extraction. We present the $\rm ^{13}CO$ narrow-velocity integrated intensity maps of these two regions in Figure~\ref{fig8}. The integrated velocity ranges are [4.5,\,5.1]  and $\rm [1.5,\,2.4] ~km~s^{-1}$, respectively. According to the morphology of $\rm 3\sigma$ contours, we identify the ridges for the striations S1-S6 in the $^{13}$CO images.

%========= Results 2
\subsection{Properties of Filamentary Structures}\label{sec:r2}
The distance of the L914 cloud listed in previous work is about $\rm 800~pc$ \citep{Dobashi1994}, which is estimated through the spectrophotometry of the nearby stars. In this work, we measure the distance of the cloud based on the MWISP CO data and $Gaia$ DR3 data \citep{Gaia2023}, using the same method as described in \citet{Yan2019}. The details about the distance measurement are presented in Appendix~\ref{sec:appendix1} and a distance of $\rm 760~pc$ is obtained to the L914 cloud. 

The optical depth of the $\rm ^{12}CO$ $J = 1-0$ line is estimated with the MWISP $\rm ^{12}CO$ and $\rm ^{13}CO$ data. Assuming that $\rm ^{12}CO$ and $\rm ^{13}CO$ have the same excitation temperature and beam filling factor, we calculate the optical depth of $\rm ^{12}CO$ as follows:
\begin{equation}
\frac{T({\rm ^{13}CO})}{T({\rm ^{12}CO)}} = \frac{1 - e^{-\tau}/R_{\rm i}}{1 - e^{-\tau}},
\end{equation} 
where $T(\rm ^{12}CO)$, $T(\rm ^{13}CO)$ and $\rm \tau$ are the peak intensity of $\rm ^{12}CO$ and $\rm ^{13}CO$ spectra, and the optical depth of $\rm ^{12}CO$, respectively. $R_{\rm i}$ represents the isotopic ratio of $\rm ^{12}CO/^{13}CO$, derived from the relation ${\rm [^{12}C/^{13}C]} = 4.08D_{\rm GC}+18.8$ \citep{Sun2024}. In this equation, $D_{\rm GC}$ denotes the Galactocentric distance and the isotopic ratio is estimated to be $R_{i} \sim 52$ for the L914 cloud. The calculated optical depth ($\tau$) varies from approximately 15 in the outskirts of the cloud to around 60 at the densest positions. The mean value of $\tau$, approximately 32, suggests that the $\rm ^{12}CO$ emission is optically thick in the L914 cloud.

We further convert the main-beam brightness temperature $T_{\rm MB}$ of $\rm ^{12}CO$ to excitation temperature $T_{\rm ex}$ under the assumption of local thermodynamic equilibrium (LTE) with the following formula:  
\begin{equation}
      T_{\rm{mb}}  = [J(T_{\rm ex}) - J(T_{\rm bg}) ][1 - e^{(-\rm\tau)}],
\end{equation}
where $T_{\rm bg} = \rm 2.7~K$ is the background temperature; $J$ is the radiation temperature and $J_{\rm \nu}(T) = T_{0}/[e^{(h\nu/k_{\rm B}T)} - 1]$, here $T_0$ is the intrinsic temperature of $^{12}$CO and $T_0 = h\nu/k_{\rm B}$, $k_{\rm B}$ and $h$ are the Boltzmann constant and Planck constant, respectively. We analyze the pixels with the integrated intensity of $\rm ^{13}CO$ larger than $\rm 1~K~km~s^{-1}~(\sim 3\sigma)$ and find that the $T_{\rm ex}$ in L914 ranges from 6 to $\rm 15~K$, with a mean value of $\rm \sim10~K$. We further assume a uniform $T_{\rm ex}$ of CO and its isotopologues, the optical depth and column density of $\rm^{13}CO$ can be calculated as follows \citep{Bourke1997}:
\begin{equation}\label{eq_tau}
{\tau_{13} = -ln[1 - \frac{T_{\rm mb}(\rm{^{13}CO})}{5.29}([e^{5.29/T_{\rm ex}} - 1]^{-1} - 0.164)^{-1}]},
\end{equation}

\begin{equation}
{N({\rm ^{13}CO}) = 2.42 \times 10^{14 }\cdot \frac{\tau(\rm ^{13}CO)}{1 - e^{-\tau (\rm ^{13}CO)}} \cdot \frac{(1+0.88/T_{\rm ex})\times \int{T_{\rm mb}({\rm ^{13}CO}) dV}}{1 - e^{-T_{0}({\rm ^{13}CO})/T_{\rm ex}}}}.
\end{equation}

The abundance ratios of $\rm {H_2}/{^{12}CO} \approx 1.1 \times 10^4$ \citep{Frerking1982} and $\rm ^{12}C/^{13}C \approx 52$ \citep{Sun2024} are used to calculate the H$_2$ column density. The column density map is displayed in Figure~\ref{fig9}. The column densities of the dense ridge and diffuse striations are $N\rm(H_{2}) \sim 5-7 \times 10^{21}~cm^{{-2}}$ and $N\rm(H_{2}) \sim 1-3 \times 10^{21}~cm^{{-2}}$, respectively. As seen in Figure~\ref{fig9}, the striations appear to be connected with the densest parts of the skeleton.

We employ the Gaussian model in python package RadFil \citep{Zucker2018} to fit the column density profiles of the L914 filament and estimate the deconvolved width ($FWHM_{\rm dec} = \sqrt{FWHM^{2} - HPBW^{2}}$, where $HPBW$ is the half-power beamwidth) of the filament. Assuming that the filamentary structure is a long cylinder, the line mass, volume density and average column density can be calculated with the equations of $M_{\rm line} = \mu m_{\rm H}\int{N(r) \it dr}$, $n = \frac{M_{\rm line}}{\pi r^{2}}$ and $\overline{N} = \frac{M_{\rm line}}{2\times r}$, where $\mu$, $m_{\rm H}$ and $r$ are the mean molecular mass, the hydrogen atom mass and the radius of the filament, respectively. With the same method, we also calculate the properties of the six striations (S1-6) identified in Section~\ref{sec:r1}. All properties of the filamentary structures are listed in Table~\ref{tab:fil}. The estimated width, line mass and volume density of the L914 filament are $\rm \sim 1~pc$, $80~M_{\odot}~\rm pc^{-1}$ and $\rm1300~cm^{-3}$, respectively. The L914 filament is thermally supercritical with the line mass much larger than the critical equilibrium value for an isothermal cylinder ($M_{\rm line, crit} \sim 17~M_{\odot}~\rm pc^{-1}$ at a gas temperature of $\rm 10~K$, derived from $M_{\rm line, crit} = 2c_{\rm s}^{2} / \rm G$; \citealt{Ostriker1964}). The mean width of the striations is about $\rm 0.5~pc$, which is half that of the L914 filament. The line masses of the striations range from 3 to $12~M_{\odot}~\rm pc^{-1}$, with a mean value of $\sim 8~M_{\odot}~\rm pc^{-1}$. All the striations are thermally subcritical. Unless additional pressure is supplied (such as magnetic compression), these striations are expected to disperse. Considering that striations are so tenuous that can be easily affected by surrounding material, large uncertainties will inevitably be introduced onto the calculations.

%========= Results 3
\subsection{Magnetic Field Traced by $Planck$ Dust Polarization}\label{sec:r3}
\subsubsection{Field-structure Orientation}\label{sec:r31}
The plane-of-sky (POS) magnetic field can be derived with the $Planck$ data based on the assumption that the short axis of dust grains is well aligned with the local direction of the magnetic field (\citealt{Andersson2015}). Figure~\ref{fig10} shows the inferred magnetic field overlaid on the MWISP $\rm^{13}CO$ zeroth-moment image. In the left part of L914, the magnetic field aligns well with the dense skeleton. However, in the central and right parts, the magnetic field appears a sharp turn, becoming perpendicular to the dense skeleton and parallel to the diffuse striation structures. In order to quantify the relative orientation between the magnetic field and filamentary structures, we first present the distribution of the polarization position angles (PAs) in Figure~\ref{fig11}, from which the PAs can be visually divided into three groups. We perform a multicomponent Gaussian function fitting on the PA distribution and obtain the mean values of $\sim29.6\arcdeg$, $-44.2\arcdeg$ and $-63.9\arcdeg$, with the 1-sigma dispersions of $\sim5.03\arcdeg$, $7.56\arcdeg$, and $4.23\arcdeg$ for the three components C1, C2, and C3, respectively. The first component C1 (purple shadow in Figure~\ref{fig11}) covers a range of PAs from $\sim9\arcdeg$ to $50\arcdeg$, and corresponds to the marked region R1 in Figure~\ref{fig10}. The second component (C2, orange shadow) ranges from $\sim -52\arcdeg$ to $-14\arcdeg$, and the corresponding area is R2. The third component (C3, green shadow) spans from $\sim -81\arcdeg$ to $-52\arcdeg$, and is located in the right of L914, i.e., region R3. The global B-field orientation, denoted as $\rm \theta_{B}$, can be derived from PA, using the formula $\rm \theta_{B} = \overline{\theta}_{PA} - \frac{\pi}{2}$.

As for the structure orientation, we first roughly break the L914 filament into three segments (Seg1-3 in Figure~\ref{fig9}) according to the distribution of dust polarization angles illustrated in Figures~\ref{fig10} and \ref{fig11}. We then employ cubic polynomial fitting to smooth the path of each filamentary structure (segments \& striations), and compute the derivative at each point along the trajectory. The mean tangential angle along each path is adopted as the structure orientation. The structure orientations of Seg1-3 also adhere to the IAU convention (see Section~\ref{sec:o2}), with orientation angles of $-88\arcdeg\pm18\arcdeg$, $-77\arcdeg\pm23\arcdeg$ and $-66\arcdeg\pm12\arcdeg$ for Seg1, Seg2, and Seg3,, respectively. Figure~\ref{fig12} presents the relative orientations of the projected magnetic fields $\rm B_{pos}$ with respect to the dense segments and diffuse striations. As seen in this image, the magnetic fields in R2 and R3 are indeed parallel to the hair-like structures, which is consistent with the definition of striations (see Section~\ref{sec:r1}). Moreover, a bimodal configuration is shown in subregions R2 and R3.

\subsubsection{Strength of Magnetic Field}\label{sec:r32}
The strength of the magnetic field projected on the plane of the sky ($B_{\rm pos}$) can be estimated with the Davis-Chandrasekhar-Fermi (DCF; \citealt{Davis1951, Chandrasekhar1953}) method. Under the assumption that the observed dispersion of polarization position angle is purely caused by incompressible and isotropic turbulence, the magnetic field strength is estimated as
\begin{equation} \label{eq2}
B_{\rm pos} = Q\sqrt{4\pi\rho}\frac{\sigma_{\rm turb}}{\sigma_{\theta}}~\rm{(cgs)},
\end{equation} 
where $\rho$ is the volume density ($\rm \rho \sim 1300~cm^{-3}$, see Section~\ref{sec:r2}), $\sigma_{\rm turb}$ is the one-dimensional non-thermal velocity dispersion, $\sigma_{\theta}$ is the dispersion in polarization angle, and $Q$ is the correction factor. The three-dimensional numerical magnetohydrodynamic simulations in \citet{Ostriker2001} suggest that the correction factor of $Q = 0.5$ should be applied to the DCF formula when the dispersion of PAs is less than $\sim 25\arcdeg$, which is applicable for our samples (see Figure~\ref{fig11}). The uncertainty of $Q$ factor is 30\% \citep{Crutcher2004}.

The PA dispersion has been derived in Section~\ref{sec:r31}. We further evaluate the nonthermal velocity dispersion ($\sigma_{\rm turb}$) with $\rm ^{13}CO$ data by eliminating the thermal portion($\rm \sigma_{th}$) from the observed velocity dispersion ($\rm \sigma_{obs}$)
\begin{equation}
\sigma_{\rm turb}^2 = \sigma_{\rm obs}^2 - \sigma_{\rm th}^2,
\end{equation}
where the thermal velocity dispersion is determined by $\sigma_{\rm th} = \sqrt{\frac{k_{\rm B}T_{\rm kin}}{m_{\rm obs}}}$, here $m_{\rm obs}$ is the mass of the $\rm ^{13}CO$ molecule ($m_{\rm obs} = \rm 29~amu$), $T_{\rm kin}$ is the gas kinematic temperature, which we adopt the excitation temperature estimated from the $\rm ^{12}CO$ emission, under the assumption of LTE (see Section~\ref{sec:r2}). The estimated median of $\sigma_{\rm th}$ is $\rm\sim 0.05\pm0.01~km~s^{-1}$ for all the three regions, which is too small and therefore negligible compared to $\sigma_{\rm obs}$. The effect of opacity ($\tau$) broadening on the velocity dispersion can be estimated by referring to (see, e.g., \citealt{Phillips1979, Hacar2016}) 
\begin{equation}\label{eq8}
\frac{\sigma_{\rm obs}}{\sigma_{\rm int}} =\frac{1}{\sqrt{\rm{ln} 2}}\left[\rm{ln} \left( \frac{\tau}{ln\left(\frac{2}{e^{-\tau}+1}\right) } \right) \right]^{1/2},
\end{equation}
where $\sigma_{\rm obs}$ and $\sigma_{\rm int}$ are the observed and intrinsic velocity dispersion, respectively. As the opacity reaches 0.6, the contribution of opacity broadening can reach up to 10\%. According to Equation~\ref{eq_tau}, we find that more than 30\% of pixels in the L914 cloud have $\rm ^{13}CO$ optical depth greater than 0.6. Therefore, the optical depth correction is necessary. In Equation~\ref{eq8}, we adopt the intensity-weighted velocity dispersion $\sigma_{\rm obs}$ to calculate $\sigma_{\rm int}$. Figure~\ref{fig13} presents the distributions of $\sigma_{\rm int}$, with the mean values of $\sim 0.6$, 0.5 and $\rm 0.4~km~s^{-1}$ for subregions R1, R2 and R3, respectively.

With Equation~\ref{eq2} and all the parameters obtained above, the magnetic field strengths are estimated to be $\rm\sim 101\pm71~\mu G$, $\rm54\pm27~\mu G$, and $\rm 75\pm45~\mu G$ for regions R1, R2, and R3, respectively (see Table~\ref{tab:b}). Since the angular resolution of $Planck$ data is $\sim 10\arcmin$, any disordered structures smaller than the beam size will be smoothed and result in an overestimate of the $B_{\rm pos}$. Also, if we use the critical density of CO $J = 1-0$ ($\rm\sim 1000~cm^{-3}$; see \citealt{Yang2010}) at the kinematic temperature of $\rm 10~K$ to calculate the $B_{\rm pos}$, the estimated strength will be 88, 47 and $\rm 66~\mu G$, respectively. Additionally, we attempt to estimate the magnetic field strength using the modified DCF method (ST method, $B_{\rm pos} = \sqrt{2\pi\rho}\frac{\sigma_{\rm v}}{\sqrt{\sigma_{\theta}}}$; see, \citealt{Skalidis2021a, Skalidis2021b}), which takes into account both the anisotropy and compressibility of turbulence. The estimated strengths of the magnetic field are 42, 28, and $\rm 29~\mu G$ in the three subregions, nearly half of the DCF results. However, as mentioned in \cite{Skalidis2021a}, the ST method will underestimate the magnetic field strength for regions where self-gravitation is non-negligible. We thus consider the results from the ST method as a lower limit of $B_{\rm pos}$.

%============================================================
%========= Discussion
\section{Discussion}

\subsection{Comparison with the Taurus and Musca Filamentary Clouds}\label{sec: d1}
The combined MWISP CO and $Planck$ dust observations show a filamentary structure in the L914 cloud, with a string of magnetically aligned striations perpendicular to the dense ridge (see, e.g., Figure~10).
Similar striation structures are also observed in the Taurus B211 (e.g., \citealt{Palmeirim2013}) and Musca filaments (e.g., \citealt{Cox2016, Bonne2020}). In comparison to the L914 cloud, these two objects are located at relatively higher Galactic latitudes and are much closer ($\rm140~pc$ for Taurus; $\rm200~pc$ for Musca), so there is very low line-of-sight contamination. The main filament of L914 spans $\rm \sim 50~pc$ in length, which is roughly five times longer than that of B211 and Musca ($\rm \sim 10~pc$). Besides, the line mass of the L914 filament ($\sim 80~\it{M}_{\odot}~\rm pc^{-1}$) is also much larger than that of the Musca and Taurus B211 filaments ($\sim20~\it{M}_{\odot}~\rm pc^{-1}$ and $50~\it{M}_{\odot}~\rm pc^{-1}$, respectively). It must be noted that the properties of filaments can vary significantly due to different tracers, sensitivities and spatial resolutions\footnote{The spacial resolution of the $Herschel$ dust observations in B211 and Musca clouds is less than $\rm 0.02~pc$, while the resolved scale of the L914 cloud in the MWISP CO observations is $\rm \sim 0.2~pc$.}. Furthermore, we reveal the quasi-periodic arrangements in both the CO intensity (see Figure~\ref{fig6}) and velocity dispersion (see Figure~\ref{fig7}), which are not observed in the cases of B211 and Musca. These newly-discovered characteristics may help us to better understand the nature of striations associated with filaments and provide new observational constraints on the theoretical models. The morphology of the B-field, revealed by polarization measurements from either starlight or dust emission, demonstrates a large-scale ordered field parallel to the striations in all three clouds. The estimated B-field strengths in Musca, Taurus and L914 clouds are $\sim \rm 12~\mu G$ \citep{Planck_2016},  $\rm 25~\mu G$ \citep{Chapman2011}, and $\rm 80~\mu G$ (this work), respectively.

\cite{Andre2014} proposed that filaments gain mass through magnetized accretion. In this scenario, B-field is dynamically dominant in the formation of filaments by channelling the material along the sub-structures onto the dense ridge. So far, there have been several cases of accretion activity identified by velocity gradients, such as OMC-1 \citep{Hacar2017}, OMC-3 \citep{Ren2021}, DR21 \citep{Cao2022}, M120.1+3.0 \citep{SunLi2023}, California \citep{Guo2021} and Serpens \citep{Gong2018, Gong2021} regions. In these cases, accretion flows converge toward the gravity center, either along the main filament or the sub-structures in the hub-filament system. In the Musca and B211 clouds, large-scale velocity gradients perpendicular to the main filaments are observed, and are considered as the evidence that the main filament is accreting material from its surroundings via striations (see \citealp{Bonne2020, Shimajiri2019, Palmeirim2013}). For L914, we check the PV profiles along the striations and find increasing velocity gradients and dispersions along the striations S5 and S6 toward the dense ridge. The PV maps are exhibited in Figure~\ref{fig14}. If gravity is responsible for the velocity gradients, we can quantitatively delineate the motion of material in the potential well. We assume that the main filament is an infinite cylinder and then use the observed line mass to estimate the free-fall velocity $v_{\rm ff}$ at a radius $r$  \citep{Palmeirim2013},
\begin{equation}\label{ff}
v_{\rm ff} = 2 \times \sqrt{GM_{\rm line}\cdot ln(\frac{D}{r}) }
\end{equation}
where D is the distance from the tail end of the striation to the dense ridge. Since we can only derive the line-of-sight (l.o.s.) velocity $V_{\rm LSR}$ and the projected position $p$, Equation~\ref{ff} can be rewritten as $v_{\rm LSR} = v_{\rm sys} + 2\sqrt{GM_{\rm line}\cdot ln(\frac{D}{p/sin\theta})}\cdot \rm cos\theta$, here $v_{\rm sys}$ is the system velocity and $\theta$ is the inclination angle of the striation against the line of sight (l.o.s.). In Figure~\ref{fig14}, we display the fitted velocity profiles under the line masses of 60, 80, and $100~M_{\odot}~\rm pc^{-1}$, respectively. The derived systematic velocities are $\sim 1.37$ and $\rm 1.52~km~s^{-1}$, respectively. The velocity reaching the surface of the dense ridge is estimated to be $\rm 2.7-3.1~km~s^{-1}$, resulting in the estimated free-fall velocity of $\rm \sim 1.3-1.7~km~s^{-1}$. With the estimated free-fall velocity, we further calculate the current mass accretion rate with the equation of $\dot{M}_{\rm line} = \rho(R) \times v_{\rm ff} \times 2\pi R$, where $\rho(R)\sim600~\rm cm^{-3}$ is the density at the radius of dense ridge $R\sim \rm 0.5~pc$. This results in the accretion rate of $\sim 170-230~M_{\odot}\rm~pc^{-1}~Myr^{-1}$. It means that the main filament of L914 would be formed in roughly $\rm 0.3-0.5~Myr$. Using a similar method, the timescales in the B211 and Musca systems are estimated to be $\rm 1-2~Myr$ \citep{Palmeirim2013} and $\rm 1~Myr$ \citep{Bonne2020}, respectively. Our results suggest that the L914 cloud is in an earlier stage of evolution compared to the B211 and Musca filaments.

To date, numerous filamentary molecular clouds have been observed in various environments (see, e.g., the review by \citealt{Hacar2023}). However, the striation associated samples are only discovered in a few of them, such as B211 (\citealp{Palmeirim2013}), Musca (\citealp{Cox2016}) and L914 in this work. Based on the comparison of striations in B211, Musca and L914, we consider the following reasons for the limited number of observed striations: 1. Striations may be transient, short-lived structures and only appear at the early formation stage of filamentary molecular clouds; 2. Striations are more diffuse and slender than the main filament. Consequently, in observations, the striations will drown in backgrounds due to a lack of strong contrast with surrounding emission or because of their small beam filling factors \citep{Heyer2016}. Therefore, large-scale surveys with both high sensitivity and spatial resolution are needed to search for more striation samples.

%============================================================
\subsection{Comparison between the Observations and Simulations}\label{sec: d2}
Similar to the field-structure orientations in R2 and R3 (see Figure~\ref{fig12}), the bimodal configurations have been revealed in multiple studies. For example, \cite{Planck_2016} statistically measured the relative orientation between gas structures and magnetic field within ten nearby Gould belt molecular clouds, using the Histogram of Relative Orientations (HRO; \citealt{Soler2013}) technique, and found that the relative orientation changes systematically with column density $N_{\rm H}$, transitioning from being parallel in the lowest density regions to perpendicular in the highest density regions. With the same method, other molecular clouds, such as Vela C \citep{Soler2017} and Serpens Main \citep{Kwon2022} have also been studied and yielded the same conclusion. The HRO method calculated the relative orientation pixel by pixel and the $Planck$ results focused on the high- and low-density medium, while \cite{LiHB2013} defined the global cloud orientation and demonstrated that the bimodal configuration exists in the $2<A_{v}<5$ medium as well (see also \citealt{Gu2019}). Such strong coupling of B-fields and filamentary structures indicates the important role that magnetic fields play in shaping the morphology of molecular clouds. In addition, we note that subregion R1 has a column density comparable to the dense ridges of R2 and R3, while the magnetic field in R1 is parallel to the gas structure. We consider that the sharp turn in the relative orientation may be caused by the stellar feedback, which compresses and reshapes both the arc-like gas and magnetic field in subregion R1 (see, e.g., \citealt{Chapman2011, Chenzw2022}). Alternatively, the parallel configuration observed in R1 may result from projection effects. For instance, if the magnetic field is primarily oriented along the line of sight but slightly tilted toward the longitudinal direction, we would observe a projected magnetic field ($B_{\rm pos}$) parallel to the filamentary structure on the plane of the sky. However, the verification of these hypotheses is beyond the scope of this work. In this work, we focus on the subregions exhibiting striations (i.e., subregions R2 \& R3).

The bimodality in the orientation between B-field and filamentary structures is also seen in synthetic polarization maps of numerical simulations (e.g. \citealt{Soler2013, Chen2016, Hennebelle2013, LiPS2019}). \cite{Soler2013} conducted a series of 3D MHD simulations with varying initial magnetic field strengths threading molecular clouds and demonstrated that the bimodal configurations are observed exclusively in the strongly magnetized clouds ($\beta = 0.1$, where $\beta$ is the squared ratio of the sound speed to $\rm Alfv\acute{e}n$ speed).

In order to evaluate whether the subregions in L914 are gravity bound or magnetically supported, we calculate the mass-to-flux ratio in units of the critical value via
\begin{equation}\label{equ:8}
\lambda_{\rm obs} = \frac{(M/\Phi)_{\rm obs}}{(M/\Phi)_{\rm cri}},
\end{equation}
where the observed mass-to-flux ratio is $(M/\Phi)_{\rm obs} = \frac{\mu m_{\rm H}N(\rm H_{2})}{B}$, and the critical value is $(M/\Phi)_{\rm cri} = \frac{1}{2\pi\sqrt{G}}$ \citep{Nakano1978}. Then, Equation~\ref{equ:8} can be simplified as described in \cite{Crutcher2004}:
\begin{equation}
\lambda_{\rm obs} = 7.6 \times 10^{-21} \frac{N(\rm H_{2})}{B_{\rm pos}},
\end{equation} 
where $N{\rm (H_2)}$ is the mean column density in units of $\rm cm^{-2}$, and $B_{\rm obs}$ is the projected B-field strength in $\rm \mu G$. The cloud region with the ratio $\lambda > 1$ is in a supercritical state and will collapse under gravity. On the contrary, the region with $\lambda < 1$ is magnetically supported. We adopt the magnetic field strengths derived from both the DCF and ST methods to calculate the ratios (see Table~\ref{tab:b}) and found that all the subregions are subcritical. According to the velocity profile fitting in Section~\ref{sec: d1}, the inclination angles 
of the striations (S5 \& S6) against the l.o.s. are estimated to be $\theta \sim 45\arcdeg$. Thus, the corrected mass-to-flux ratio should be $\lambda_{\rm obs} \cdot \rm cos\theta$. This correction reinforces our conclusion that the L914 cloud is in a magnetically subcritical state.

As a key parameter to describe the relative importance of turbulence and magnetic field, the $\rm Alfv\acute{e}nic$ Mach number is also calculated with the following formula:
\begin{equation}
M_{\rm A, 3D} =  \frac{\sqrt{3} \sigma_{\rm NT}}{V_{\rm A, 3D}},
\end{equation}
where $V_{\rm A, 3D} = \frac{B}{\sqrt{\mu_{0}\rho}}$ is the 3D $\rm Alfv\acute{e}n$ velocity. $\mu_{0}$ is the permeability of vacuum and $B = \frac{4}{\pi} B_{\rm pos} $ \citep{Crutcher2004} is the total magnetic field strength. The $\rm Alfv\acute{e}nic$ Mach numbers of the three subregions are all less than 1 (see Table~\ref{tab:b}), indicating that the L914 cloud is sub-$\rm Alfv\acute{e}nic$.

These results demonstrate a significant dominance of the magnetic field over gravity or turbulence in the L914 cloud. In this case, the magnetic field introduces anisotropies into the cloud, assisting in the formation of filamentary structures, since motions perpendicular to the field lines are restricted by the Lorentz force \citep{Soler2017}. Conversely, in the case of super-$\rm Alfv\acute{e}nic$ turbulence ($M_{\rm A} > 1$), motions are expected to be more random. The MHD and hydrodynamical simulations presented in \cite{Hennebelle2013} also suggest that the magnetic field could increase the ellipticity of the clumps, making them more filamentary.

\cite{Tritsis2016} conducted the Kelvin-Helmholtz instability and MHD simulations to reproduce the Taurus striations\footnote{To avoid confusion, the Taurus striations referred to here are different from the B211 striations \citep{Palmeirim2013}. The former is not associated with any filaments.} \citep{Goldsmith2008}. The contrast of the intensity on and off the Taurus striations is $\sim 25\%$ and \cite{Tritsis2016} suggested that only the non-linear coupling of MHD waves can reproduce the density contrast. The contrasts in the subregions R2 and R3 of the L914 cloud are roughly 12\% and 28\%, respectively (see Section~\ref{sec:r1}). We further estimate the magnetic Jeans length ($\rm \lambda_{J, mag}$) of the L914 filament using the following equation \citep{Krumholz2019}:
\begin{equation}
\lambda_{\rm J, mag} = \lambda_{\rm J} ~(1+\beta^{-1})^{\frac{1}{2}} = \left[  \frac{\pi c_{\rm s}^{2}(1 + \beta ^{-1})} {G\rho} \right]^{\frac{1}{2}},
\end{equation}
where $\rho$ and $\lambda_{\rm J}$ represent the mass density and the standard Jeans length including only thermal pressure, respectively. The plasma beta ($\beta$) is calculated as $\beta = \frac{c_{\rm s}^2}{V_{\rm A, 3D}^2}$. Using the parameters obtained above, we estimate the magnetic Jeans length to be in the range of $\sim 2$ to $\rm 10~pc$. However, the $\rm H_2$ number densities of the striations tend to be much lower than those in the filaments (i.e., R1, R2, R3). According to the relation that magnetic field strengths are proportional to the $\rm H_{2}$ number densities \citep{Crutcher2019}, the magnetic field strengths of the striations are estimated to be much weaker. For instance, if we take $ B = 11~\mu\rm G$, the plasma parameter $\beta$ becomes 0.1, a value commonly adopted in the high magnetization model \citep{Soler2013}. Consequently, the estimated magnetic Jeans length is approximately $\rm 1.4-1.8~pc$, which is consistent with the period of the striations observed in this work (see Section~\ref{sec:r1}). Our results suggest that the striations are likely produced by MHD processes. If the striations are created by fast magnetosonic waves, \cite{Tritsis2018} predicted that the power spectra of column density cuts perpendicular to striations have peaks at roughly the same wavenumbers with the velocity centroid power spectra, and proposed a new method to calculate the magnetic field strength,
\begin{equation}
B_{\rm pos} =  \Gamma_n N_{\rm H_2}\sqrt{4\pi\rho},
\end{equation}
where $\Gamma_n$ is the square root of the ratio of the power in the velocity power spectrum and the power in the column density power spectrum,  $\Gamma_n = \sqrt{\frac{|P_{v1}|^2}{|P_{N_{\rm H_2}^1}|^2}}$. The mean column densities $N_{\rm H_2}$ and the mean density $\rho$ can be computed from the $\rm \overline{N}$ and $n$ of the striations listed in Table~\ref{tab:fil}. We present the power spectra in subregions R2 \& R3 in Figure~\ref{fig15} and \ref{fig16}a, respectively. The correlation of the periodicity of the velocity and the density exists only in subregion R3 (not in subregion R2). Based on this correlation, we calculate a magnetic field strength of $\rm47\pm12~\mu G$, which falls within the range of the results obtained by the DCF and ST methods.

%Further investigations with higher-resolution observations and additional numerical simulations are required to test it.
%column density maps derived from the $\rm ^{12}CO$ emission, $N_{\rm H_2} = X \int{T_{\rm MB, ^{12}CO}dV}$, where $X = {\rm 2.0 \times 10^{20}~cm^{-2}~K^{-1}~km^{-1}~s}$ \citep{Bolatto2013}.

Combining the identified accretion activities in Section~\ref{sec: d1}, we suggest a simple scenario where, in the early stage, a sheet-like cloud with the magnetic field parallel to it fragments into filaments under the Jeans instability. The self-gravitating filaments, perpendicular to the magnetic field, continuously accrete material from their surroundings along the magnetic field lines (through striations observed in the sensitive observations). When the filaments gain sufficient material, axial fragmentation then occurs in the MHD process with a characteristic scale corresponding to the magnetically Jeans length (i.e., the period observed in this work). 

%============================================================
%========= Summary
\section{Summary}
We present the MWISP CO ($J = 1-0$) multi-line observations toward the L914 cloud, using the PMO 13.7 \,m millimeter telescope. We reveal filamentary structures (dense skeleton \& diffuse striations) in the L914 cloud and derive their properties from the CO data. Combined with the dust polarization data from the $Planck$ survey, we further investigate the relationship between B-field and filamentary structures, aiming to understand the formation mechanism of filaments and the role of B-fields in this process. The main results are summarized below.

\begin{enumerate}
\item L914 is a filamentary molecular cloud located at a distance of $\rm \sim 760~pc$. Based on the $\rm C^{18}O$ data, we identify the dense ridge of the L914 filament, using the DisPerSE algorithm. The length, width and line mass of the L914 filament are $\rm \sim 50~pc$, $\rm1~pc$ and $\sim 80~M_{\odot}~\rm pc^{-1}$, respectively. The L914 filament is thermally supercritical with the line mass much larger than the critical line mass ($M_{\rm line, crit} \sim 17~M_{\odot}~\rm pc^{-1}$).

\item A group of hair-like striations are discovered in the two subregions of the L914 filamentary cloud. These striations are connected to the dense ridge of the filament and display quasi-periodic characteristics with an oscillation period of about $\rm 1.1-1.4~pc$. For the striations with prominent $\rm ^{13}CO$ emission (S1-S6), we estimate their basic physical properties. The mean width and line mass of these striations are $\sim \rm 0.5~pc$ and $8~M_{\odot}~\rm pc^{-1}$, respectively. All striations are in a thermally subcritical state.

\item The PV diagrams along two of the striations (S5 and S6) present increasing velocity gradients and velocity dispersions toward the dense ridge, which could be well fitted by the free-fall motions, suggesting that material flows along the striations toward the dense ridge under gravity. 

\item The magnetic field inferred from $Planck$ $\rm 353~GHz$ observation presents a bimodal configuration. The large-scale ordered B-field is well aligned with the diffuse striations, but perpendicular to the dense ridge. We estimate the strength of the B-field and evaluate the relative importance between the gravity, turbulence and B field, and suggest that B-fields play an important role in the formation of filaments by channelling the material along the striations onto the dense ridge.

\end{enumerate}

%============================================================
%========= Acknowledgements
\acknowledgments
We thank the anonymous referee for providing insightful suggestions and comments, which helped us to improve this work. This research made use of the data from the Milky Way Imaging Scroll Painting (MWISP) project, which is a multiline survey in $\rm ^{12}CO$/$\rm ^{13}CO$/$\rm C^{18}O$ along the northern Galactic plane with the PMO 13.7 m telescope. We are grateful to all the members of the MWISP working group, particularly the staff members at the PMO 13.7 m telescope, for their long-term support. MWISP was sponsored by the National Key R\&D Program of China with grants 2023YFA1608000, 2017YFA0402701, and the CAS Key Research Program of Frontier Sciences with grant QYZDJ-SSW-SLH047. This work is supported by the National Natural Science Foundation of China (grant No. 12041305). X. Chen acknowledges the support by the CAS International Cooperation Program (grant No. 114332KYSB20190009).

%============================================================
%========= Bibliography
\bibliographystyle{aasjournal}
\bibliography{ms}

%============================================================
%========= Figure

\begin{figure}
\hspace{-0.5cm}
\includegraphics[scale=0.42]{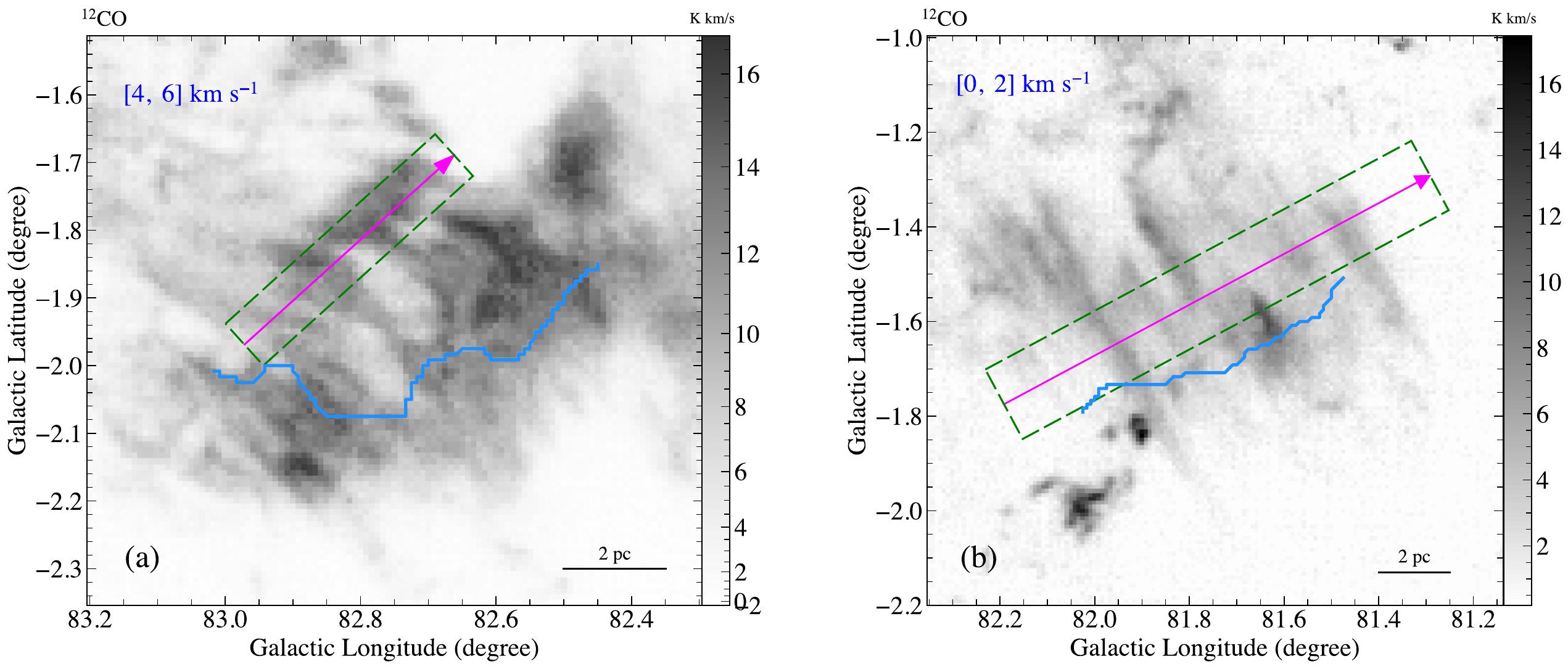}
\caption{$\rm ^{12}CO$ striations revealed in different velocity channels. The striations found in the $\rm [4,\, 6]~km~s^{-1}$ velocity range are located in the central part of the L914 cloud, while the striations in $\rm [0, \,2]~km~s^{-1}$ are situated in the western region. The blue solid line shows part of the dense ridge related to the striations, which is identified with the DisPerSE algorithm. The green rectangle demonstrates the slice used for intensity profile plotting (see Figure~\ref{fig6}). The magenta arrow line shows the path used for PV map (see Figure~\ref{fig7}).}
\label{fig5}
\end{figure}

\begin{figure}
\hspace{-0.5cm}
\includegraphics[scale=0.4]{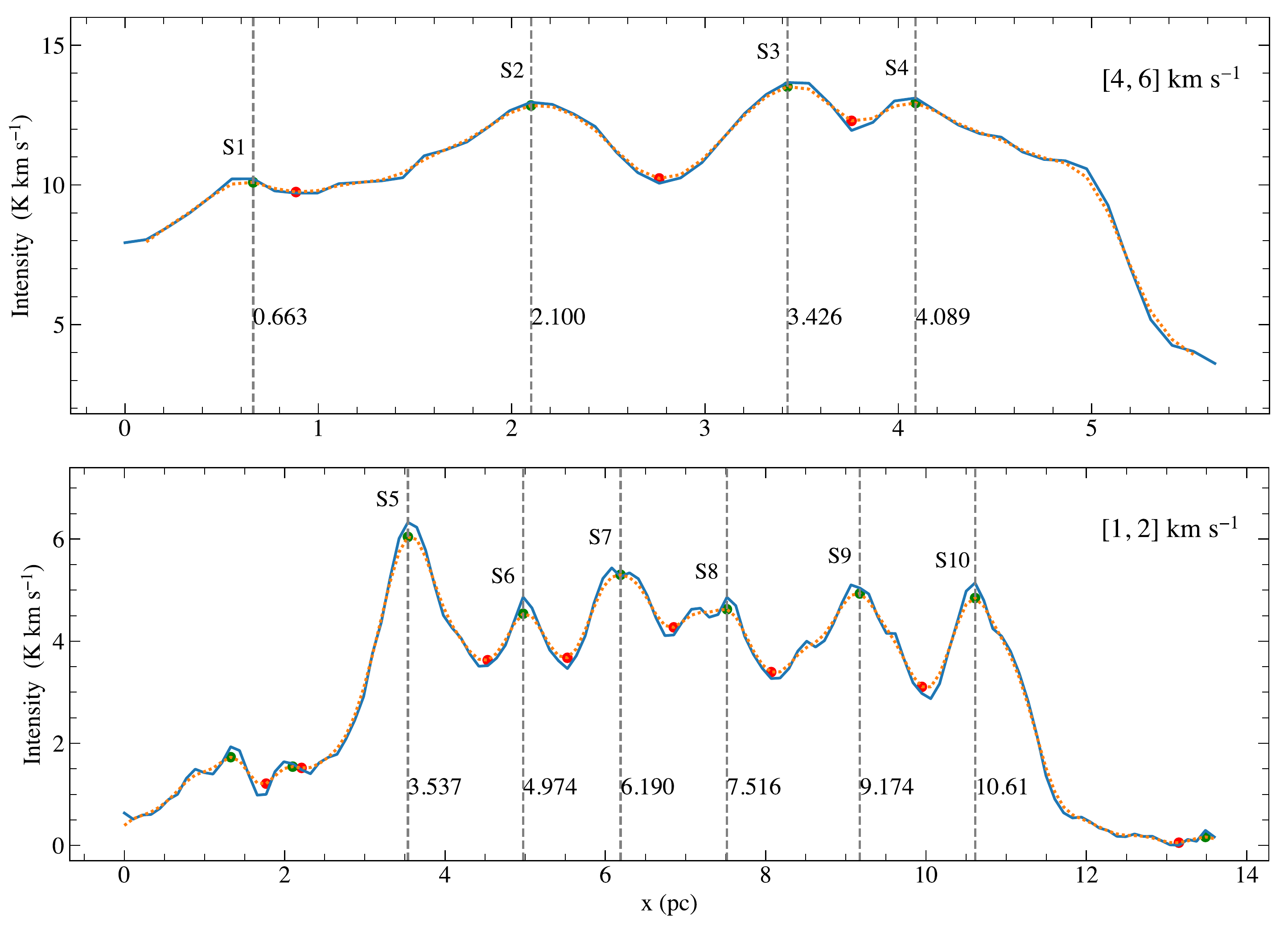}
\caption{Profiles (blue solid lines) of averaged intensity along the direction perpendicular to the striations (magenta arrow lines in Figure~\ref{fig5}). The orange dotted line shows the profile after applying a low-pass noise filter. The green and red dots represent the local maxima and minima, respectively.}
\label{fig6}
\end{figure}

\begin{figure}
\centering
\includegraphics[scale=0.35]{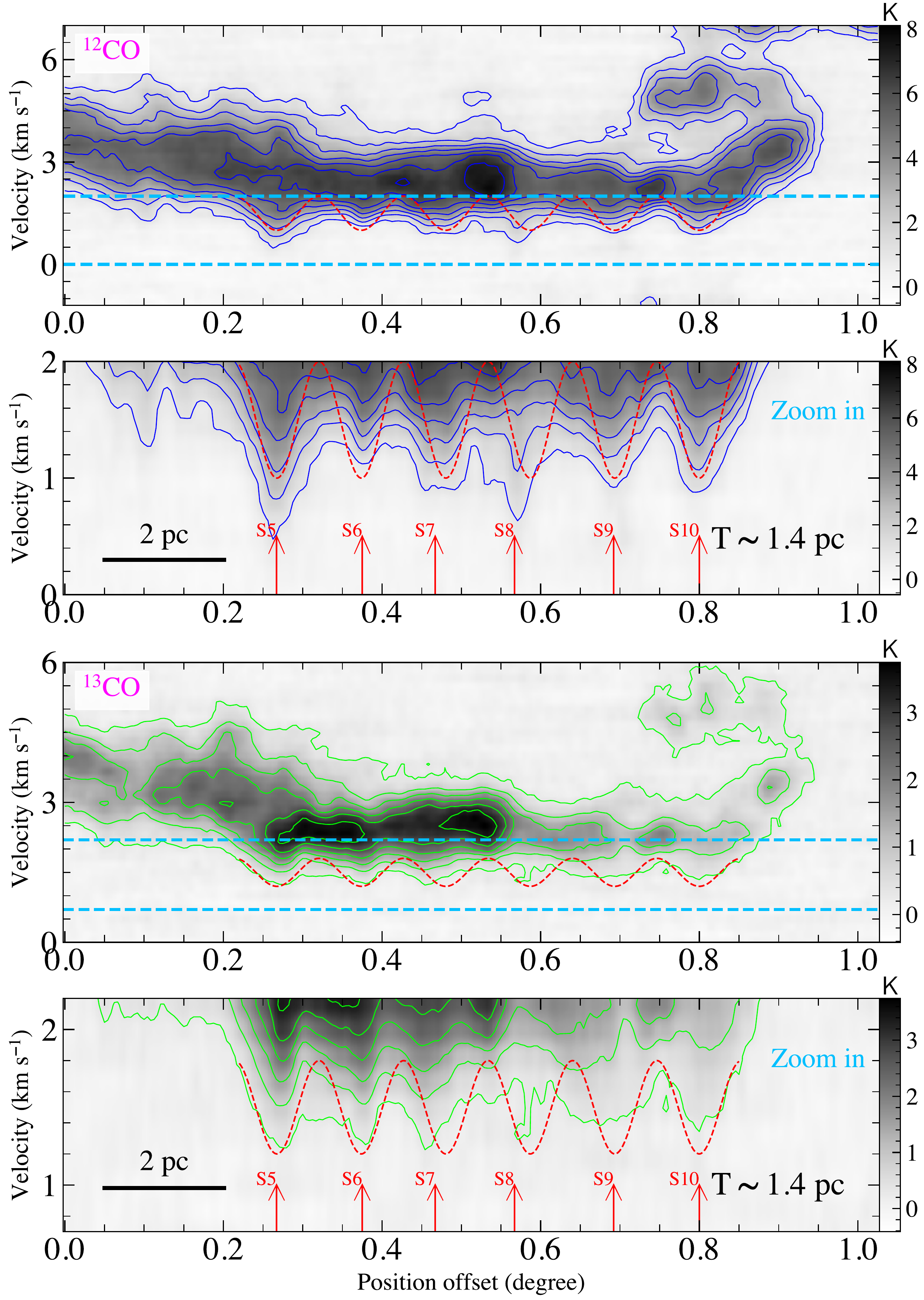}
\caption{PV maps along the arrow line shown in Figure~\ref{fig5}b. The top two panels depict the $\rm ^{12}CO$ emission, while the bottom two panels display the $\rm ^{13}CO$ emission. The $\rm ^{12}CO$ contours start at $\rm 1~K$ and increase in steps of $\rm 1~K$ ($\rm \sigma \sim 0.3~K$). The $\rm ^{13}CO$ contours start at $\rm 0.3~K$ with an increasing step of $\rm 0.6~K$ ($\rm \sigma \sim 0.1~K$). The blue dashed lines in the first and third panels indicate the velocity ranges, which are magnified in the second and fourth panels, respectively. The red dashed line shows the 'by-eye' fitting of sine waves to oscillating velocity dispersion, with the oscillation period of $\rm T\sim 1.4~pc$. The vertical arrow lines mark the positions of striations according to the local peaks of the $\rm ^{12}CO$ emission (see Figure~\ref{fig6}).}
\label{fig7}
\end{figure}

\begin{figure}
\hspace{-0.6cm}
\includegraphics[scale=0.45]{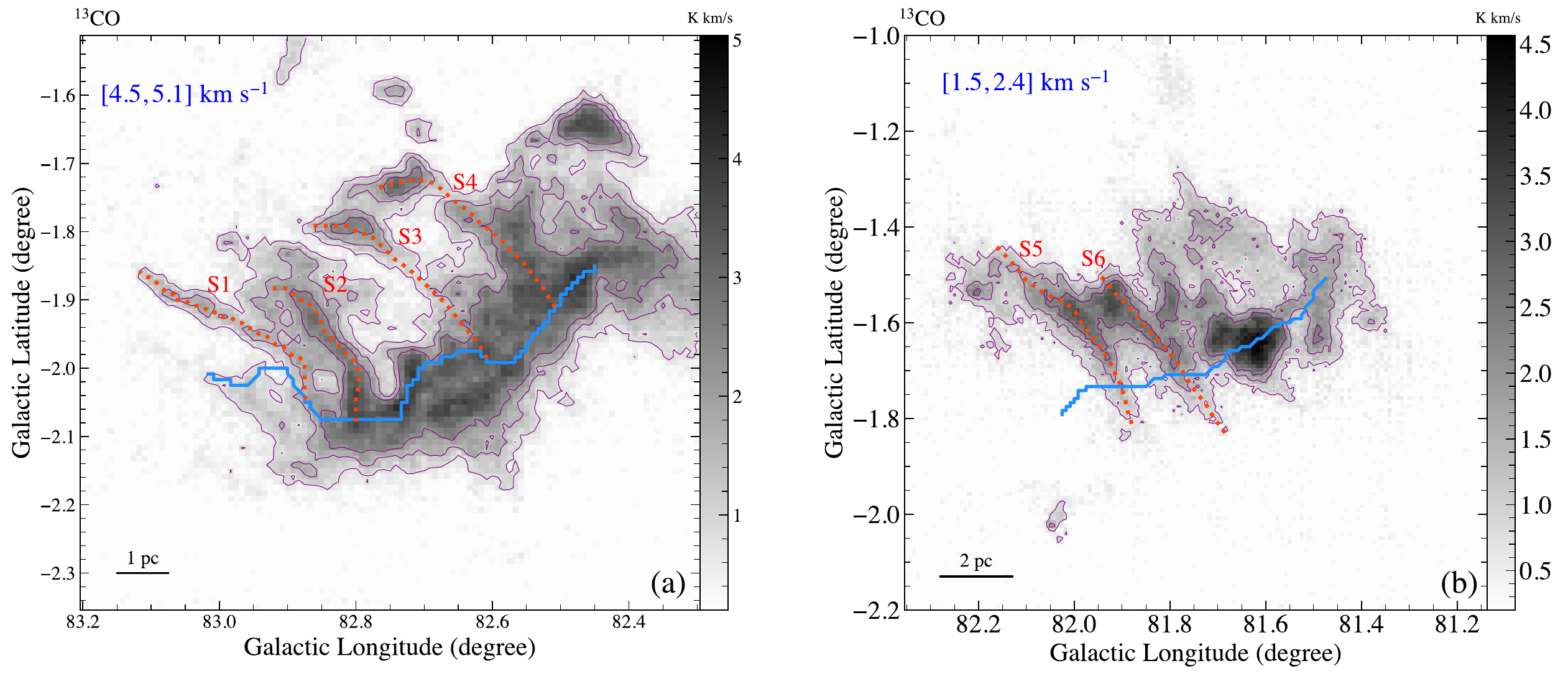}
\caption{$\rm ^{13}CO$ narrow-velocity integral maps of the striations. The gray-scale backgrounds represent the integrated intensity of $\rm ^{13}CO$ emission with the integrated velocity ranges of [4.5, 5.1] and $\rm [1.5, 2.4]~km~s^{-1}$, respectively. The contour levels are at 0.75, 1.5 and $\rm 2.25~K~km~s^{-1}$ ($\rm \sigma \sim 0.25~K~km~s^{-1}$). Striations S1-S6 (see Section~\ref{sec:r1}) are shown with red dotted lines. The blue solid line represents the dense ridge (the same as Figure~\ref{fig5}).}
\label{fig8}
\end{figure}

\begin{figure}
\centering
\includegraphics[scale=0.45]{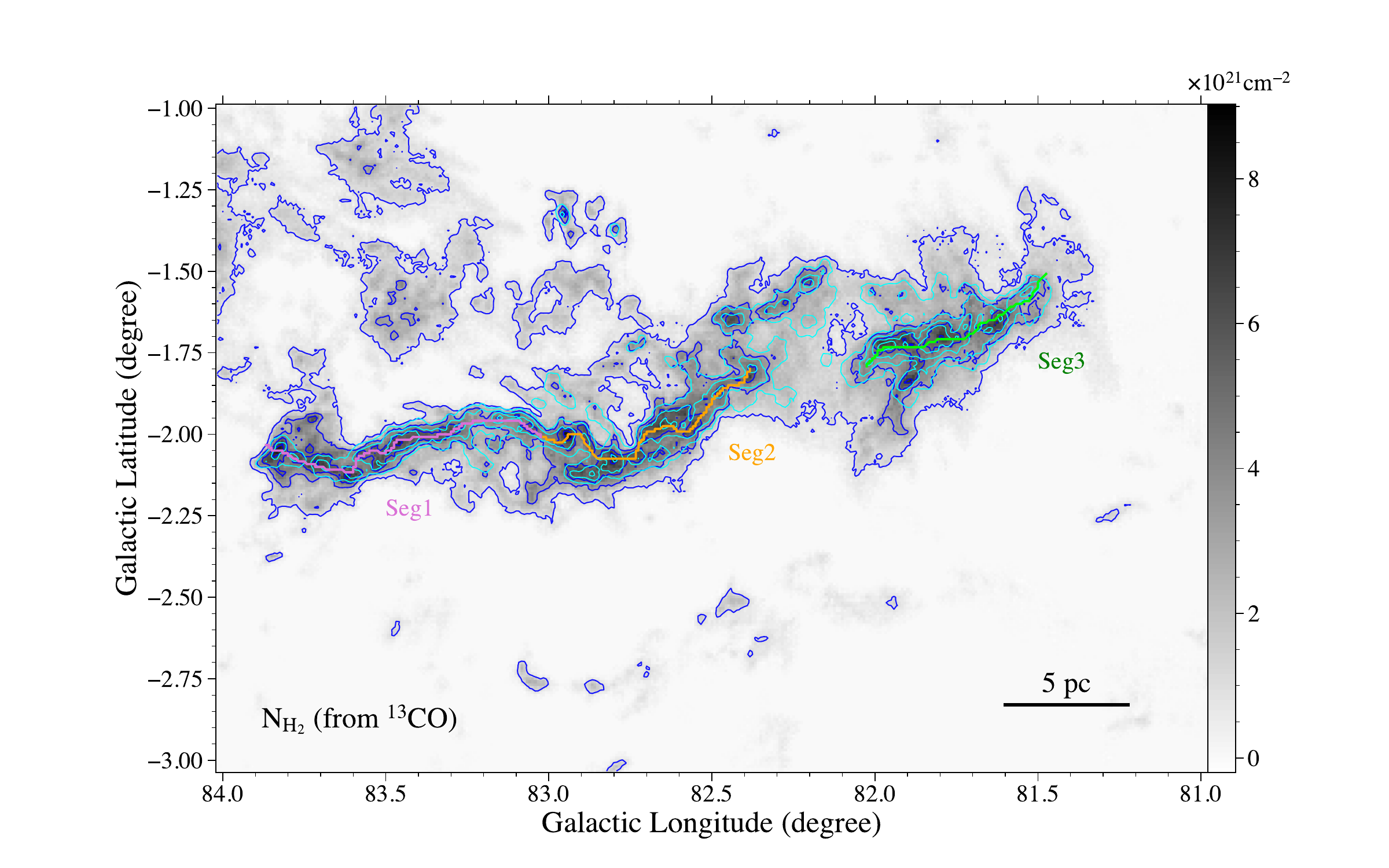}
\caption{$\rm H_{2}$ column density map calculated with the $\rm ^{13}CO$ emission. The blue contours show the density levels of 1, 3, 5, $\rm 7~\times~10^{21}~cm^{-2}$. The cyan contours show the $\rm C^{18}O$ integrated intensity with the levels at 0.3, 0.9, 1.5 and $\rm 2.1~K~km~s^{-1}$. The purple, orange, and lime lines represent the segments in Section~\ref{sec:r31}.}
\label{fig9}
\end{figure}

\begin{figure}
\centering
\includegraphics[scale=0.6]{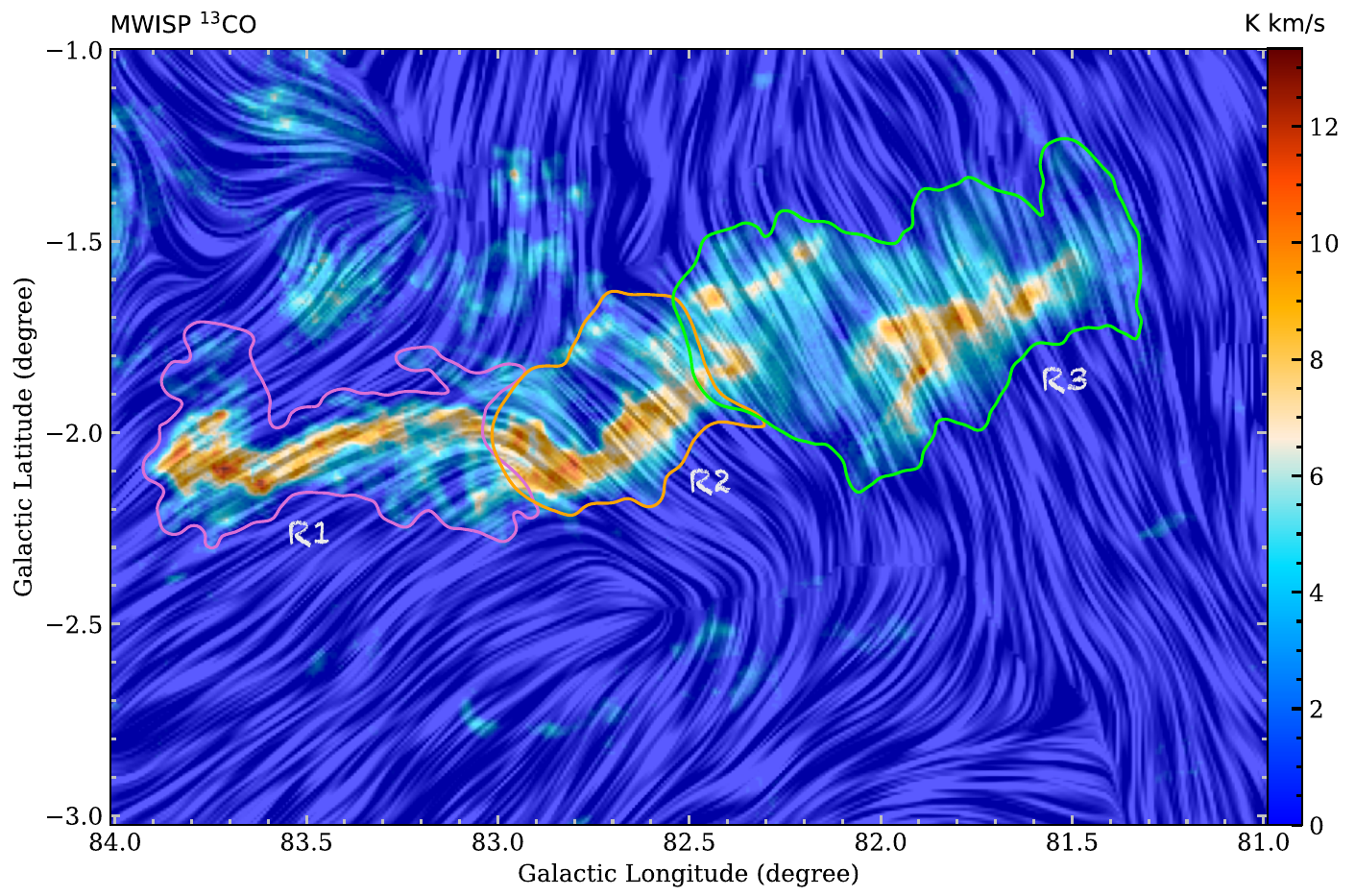}
\caption{Plane-of-sky magnetic field overlaid on the MWISP $\rm ^{13}CO$ integrated intensity map. The colorful background shows the $\rm ^{13}CO$ emission, integrated over the velocity range of $\rm0 - 6~km~s^{-1}$. The ``drapery" pattern, produced with the line integral convolution (LIC) method \citep{Cabral1993}, indicates the orientation of magnetic field lines derived from the $Planck$ dust polarization data. The colorful contours show three subregions with different polarization angles (see Figure~\ref{fig11}).}
\label{fig10}
\end{figure}

\begin{figure}
\centering
\includegraphics[scale=0.49]{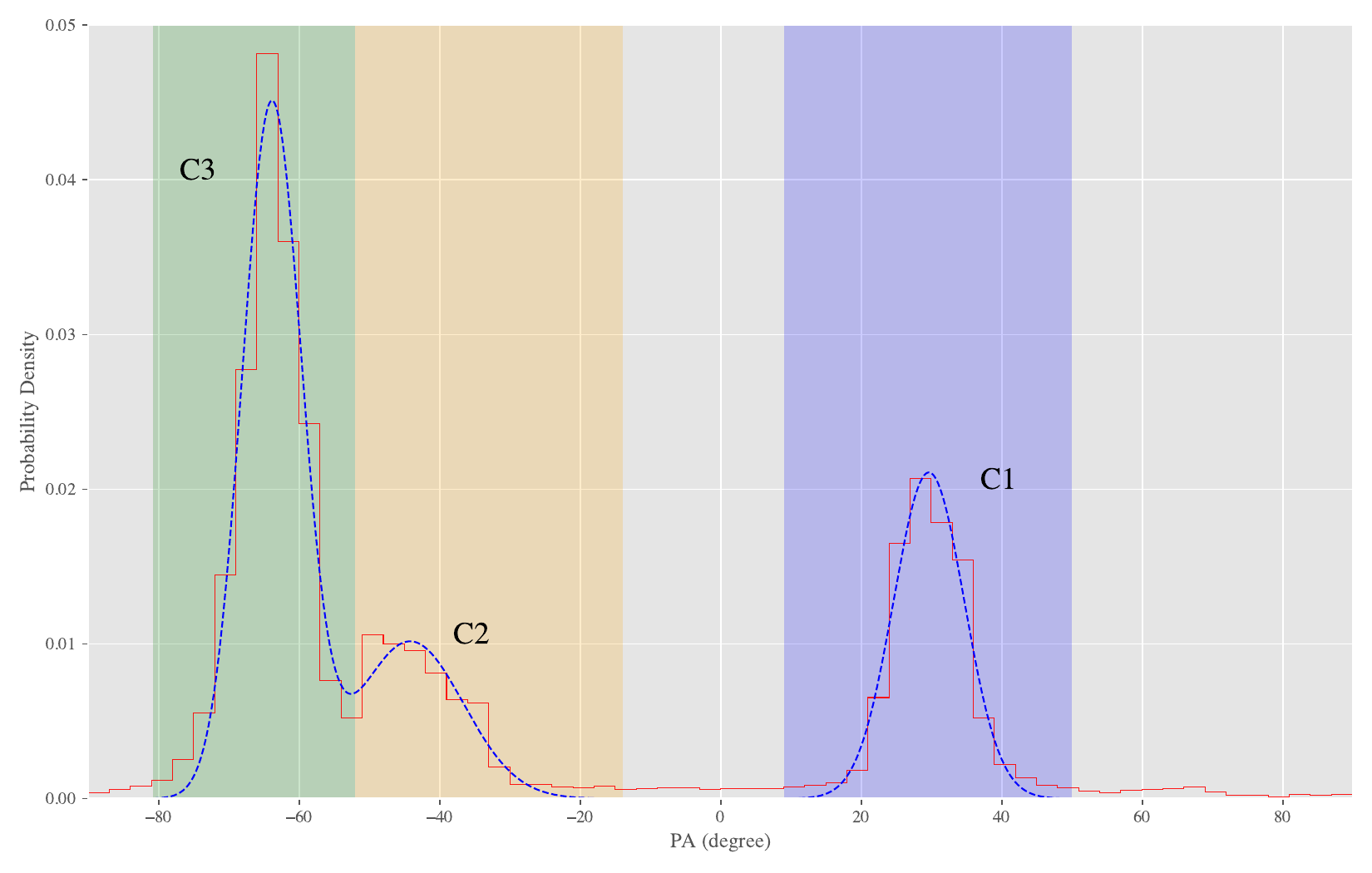}
\caption{Statistics of polarization position angles (red step line). The blue dashed line shows the fitting result with a multicomponent Gaussian function. The green, orange, and blue shadow areas represent the angle ranges of the three components, respectively.}
\label{fig11}
\end{figure}

\begin{figure}
\centering
\includegraphics[scale=0.5]{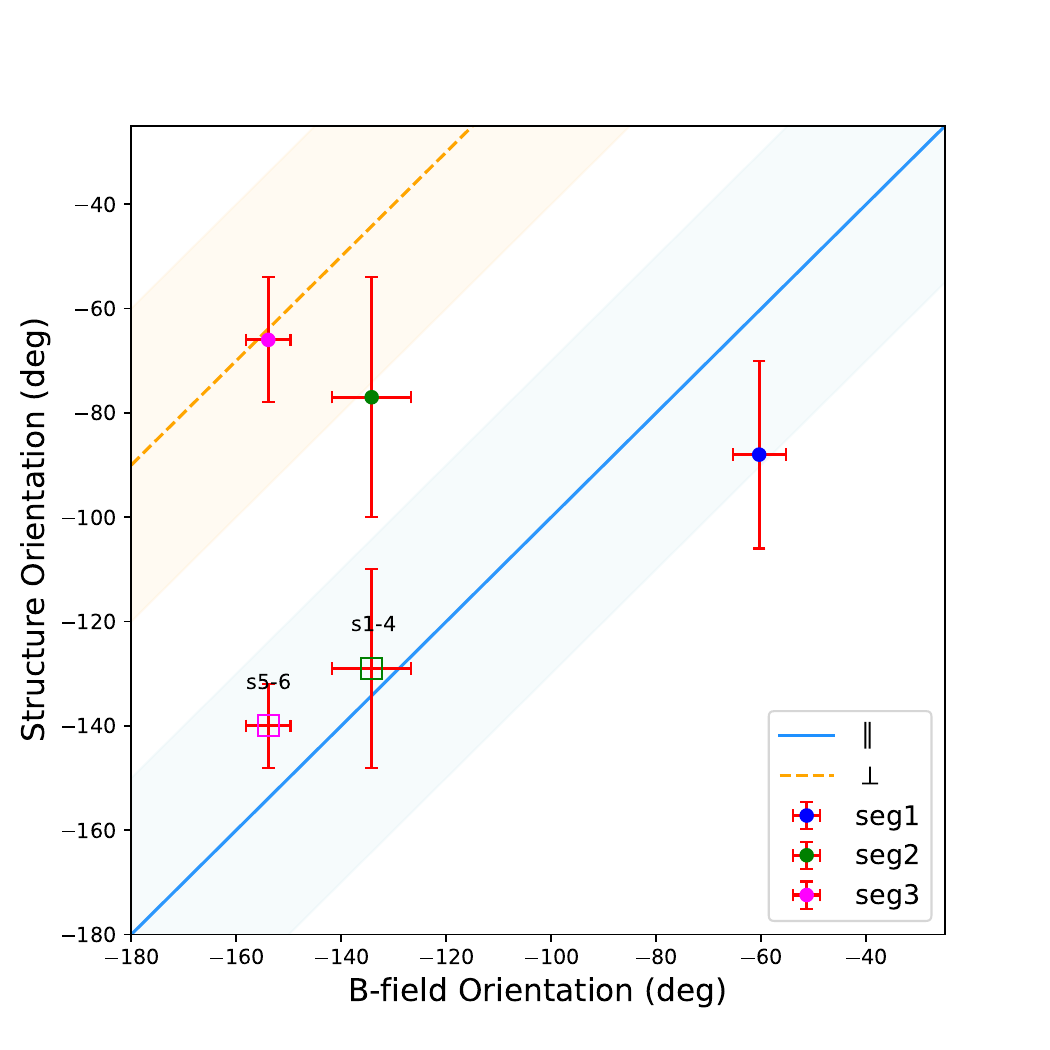}
\caption{Relative angles between B-fields and filamentary structures. The colorful dots show the relative orientation of segments and large-scale B-fields, while the distribution of striations is represented in colorful squares. The blue and orange shadow areas represent the regions that are within $30\arcdeg$ from parallelism and perpendicularity, respectively.}
\label{fig12}
\end{figure}

\begin{figure}
\centering
\includegraphics[scale=0.4]{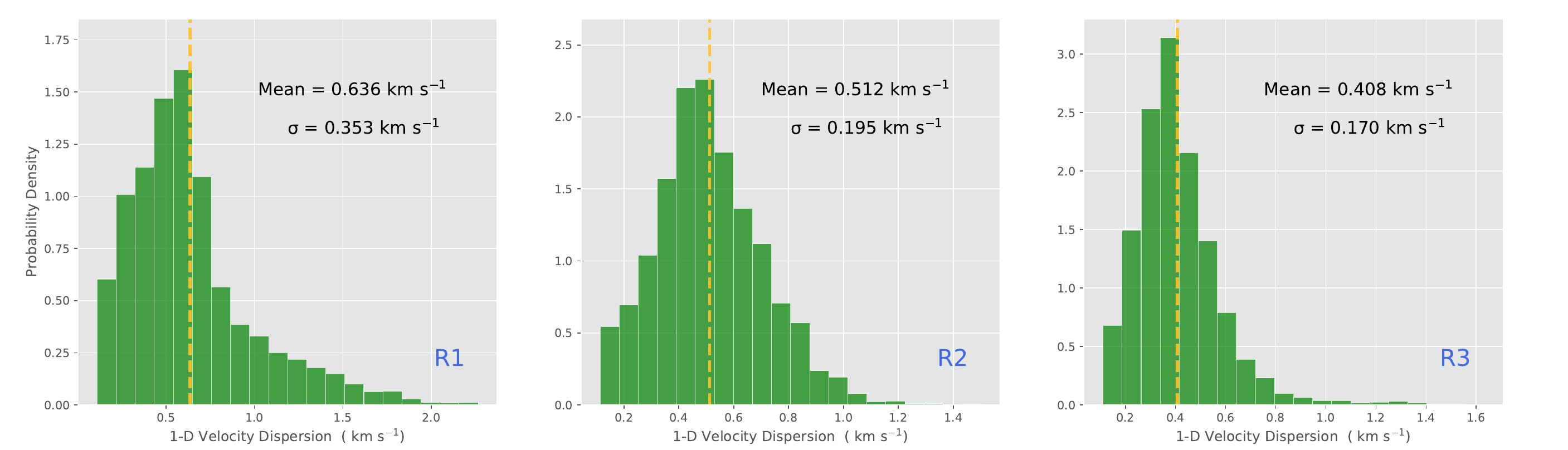}
\caption{Distribution of 1-D velocity dispersion in the three subregions.}
\label{fig13}
\end{figure}

\begin{figure*}
\gridline{\fig{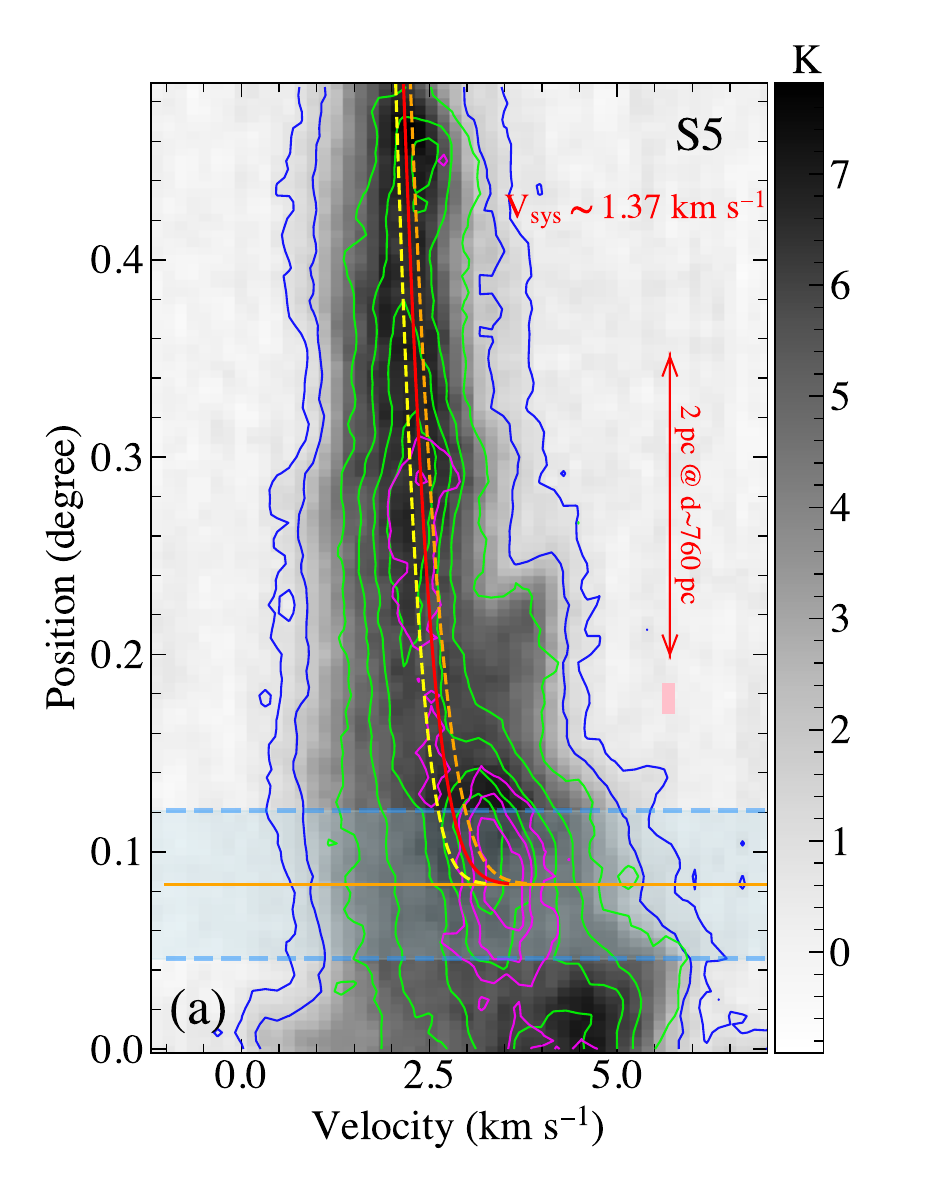}{0.5\textwidth}{ }
              \fig{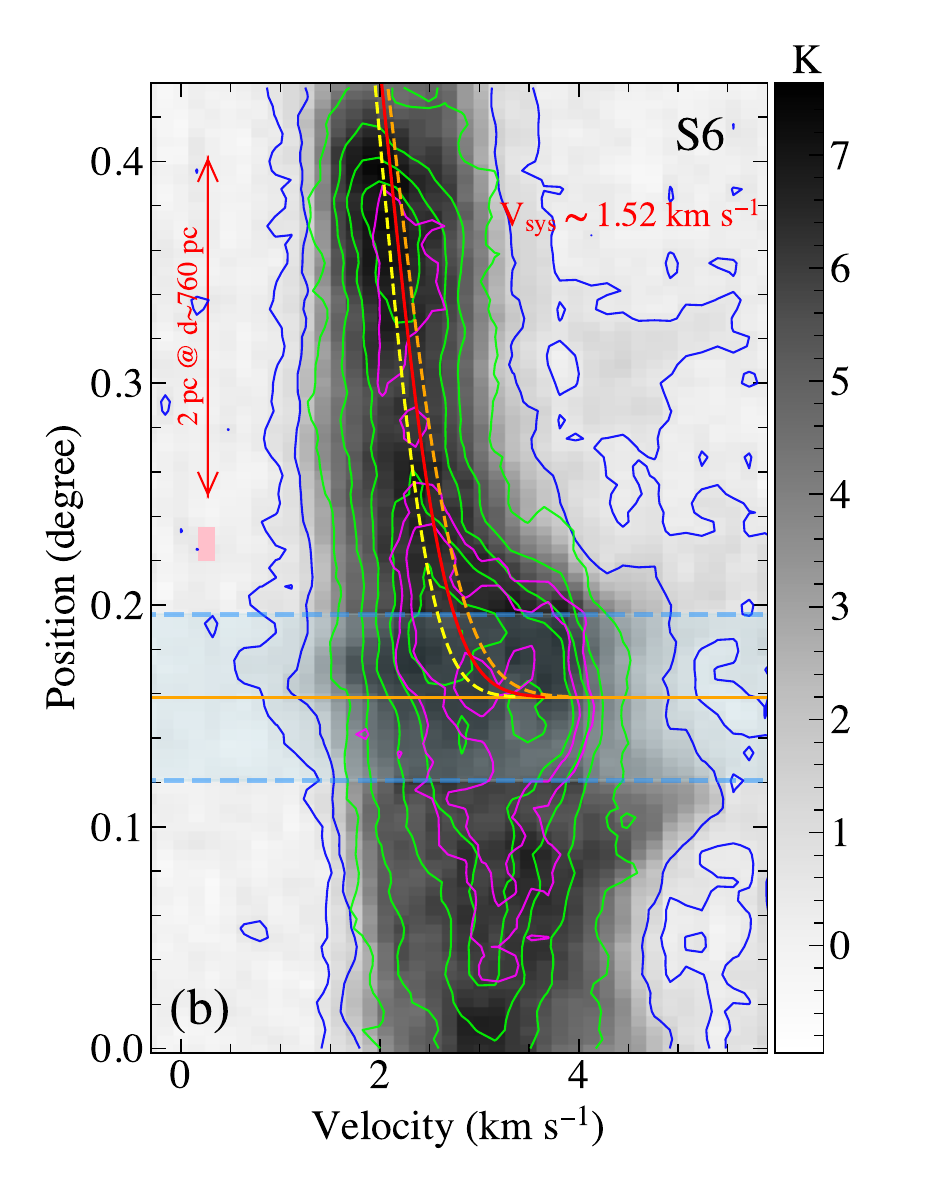}{0.5\textwidth}{ }
              }
\caption{PV maps along the striations S5 and S6. The gray-scale background shows the $\rm ^{12}CO$ emission, and the blue contours are at 0.5 and $\rm 1.5~K$ ($\rm \sigma_{12} \sim 0.2~K$). The overlapped lime and magenta contours show the $\rm ^{13}CO$ and $\rm C^{18}O$ emission, respectively. The lime contours start at $\rm 0.5~K$ and increase with the step of $\rm 1~K$ ($\rm \sigma_{13} \sim 0.15~K$). The magenta contours start at $\rm 0.5~K$ and increase with the step of $\rm 0.5~K$ ($\rm \sigma_{18} \sim 0.15~K$). The orange solid line represents the dense ridge and the blue shadow region shows its width. The red solid curve is the fitted velocity profile for a free-falling particle under the gravity of a cylinder with the line mass of $80~M_{\odot}~\rm pc^{-1}$ (see Section~\ref{sec: d1} for more details). The two dashed lines display the expected velocity profiles by changing the line masses to $60$ (yellow) and $100~M_{\odot}~\rm pc^{-1}$ (orange). The scaling is shown with the vertical arrow line and the resolution is marked with the pink rectangle.}
\label{fig14}
\end{figure*}

\begin{figure}
\centering
\includegraphics[scale=0.56]{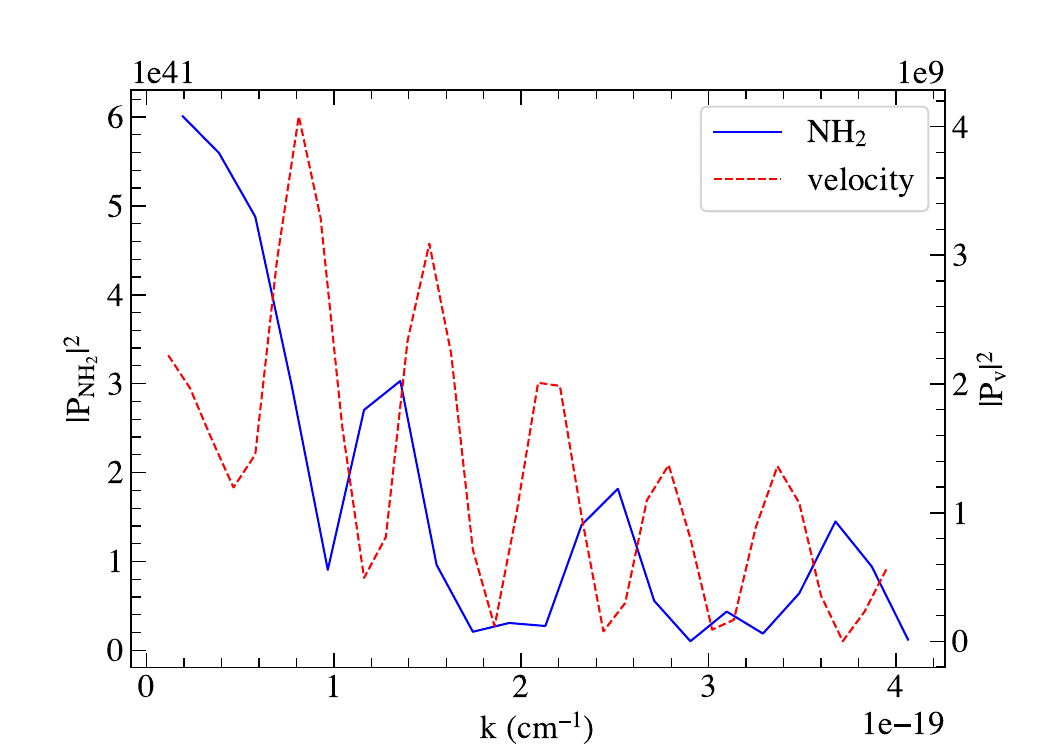}
\caption{Power spectra of column density (blue lines) cuts perpendicular to striations in subregion R2 and velocity centroid power spectra (red dashed lines). The column density is derived from the $\rm ^{12}CO$ emission with the equation of $N_{\rm H_2} = X \int{T_{\rm MB, ^{12}CO}dV}$, where $X = {\rm 2.0 \times 10^{20}~cm^{-2}~K^{-1}~km^{-1}~s}$ \citep{Bolatto2013}.}
\label{fig15}
\end{figure}

\begin{figure}
\centering
\includegraphics[scale=0.56]{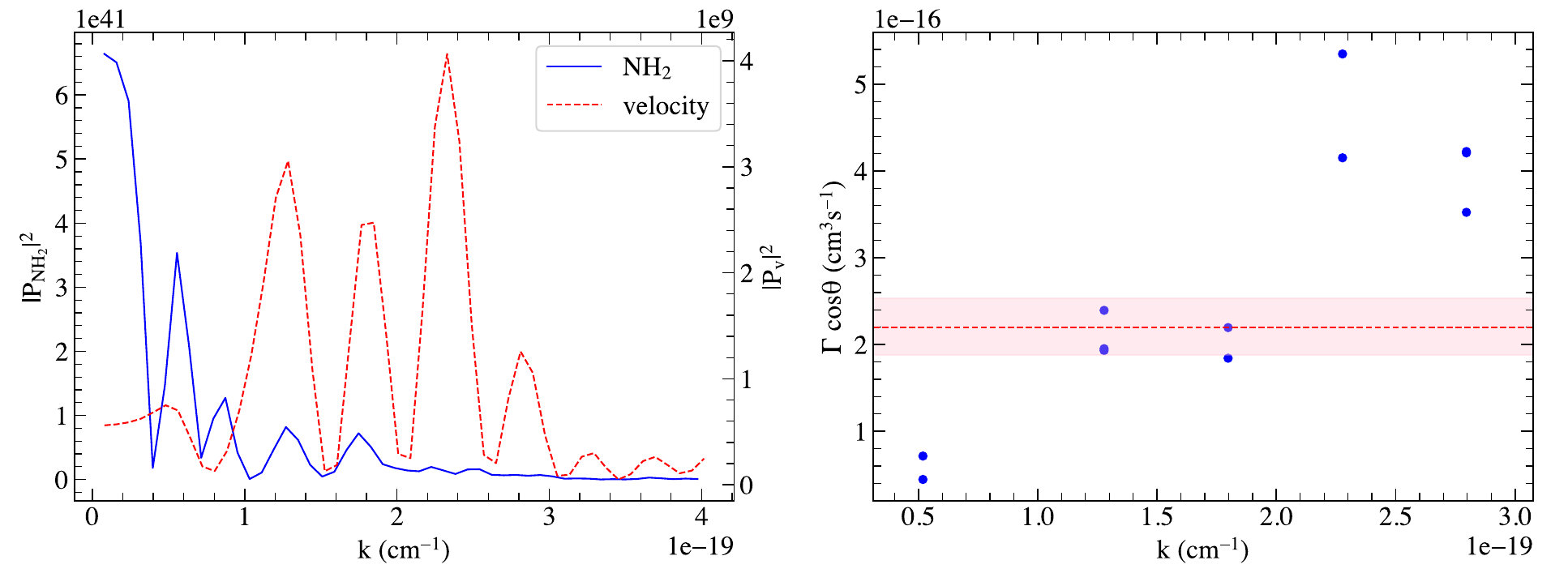}
\caption{Left panel: Same as Figure~\ref{fig15} but for subregion R3. Right panel: Distribution of parameter $\Gamma$ as a function of the wavenumber for peaks in the power spectra shown in the left panel. The red dashed line shows the fitting result obtained with the orthogonal distance regression (ODR) method, and the shaded region shows the $1\sigma$ error of the fit.}
\label{fig16}
\end{figure}

\clearpage

%============================================================
%========= Table 1
\begin{longtable}{ccccccccc}
\caption{Properties of Filamentary Structures  \label{tab:fil}}\\\hline\hline

\centering
Name & $N_{\rm peak}$ & Width & Length & $M_{\rm line}$ & n & $\rm \overline{N}$ & $M_{\rm tot}$ & Orientation \\
           & ($\rm 10^{21}~cm^{-2}$) & (pc) & (pc) & ($M_{\odot}~\rm pc^{-1}$) & ($\rm cm^{-3}$) & ($\rm 10^{21}~cm^{-2}$) & ($M_{\odot}$) & (degree) \\\hline 
\endfirsthead
\hline
\endfoot
\hline\\
\multicolumn{9}{c}{\parbox{\textwidth}{\textbf{Note. }Properties of the filamentary structures. Column 1 lists the names of the L914 filament and striations. Columns 2 and 3 are the peak column density and width obtained from the Gaussian fitting. Column 4 is the length of the structure. Columns 5--8 are the line mass, volume density, averaged column density and total mass. Column 9 lists the orientation angle of the filamentary structure, measured counterclockwisely from the Galactic north in degrees.}}
\endlastfoot
Fil & $3.94\pm0.3$  & $1.06\pm0.2$ & $48.5\pm0.1$ & $77\pm16$ & $1300\pm300$ & $3.20\pm1.0$ &  $3700\pm800$  &  - \\
S1 & $0.76\pm0.1$ & $0.44\pm0.1$ & $4.6\pm0.1$ & $6\pm2$   & $560\pm200$ & $0.60\pm0.1$ & $28\pm9$   & $-120\pm12$ \\
S2 & $1.34\pm0.1$ & $0.50\pm0.2$ & $3.7\pm0.1$ & $12\pm5$ & $870\pm360$ & $1.06\pm0.1$ & $44\pm19$   & $-138\pm22$ \\
S3 & $0.47\pm0.1$ & $0.33\pm0.2$ & $4.7\pm0.1$ & $3\pm2$   & $500\pm360$   & $0.40\pm0.1$ & $14\pm9$   & $-129\pm18$ \\
S4 & $0.80\pm0.2$ & $0.38\pm0.2$ & $4.6\pm0.1$ & $6\pm4$ & $760\pm400$ & $0.70\pm0.2$ & $28\pm19$   & $-129\pm23$ \\
S5 & $0.99\pm0.1$ & $0.68\pm0.2$ & $6.6\pm0.1$ & $12\pm4$ & $500\pm220$   & $0.78\pm0.1$ & $79\pm26$ & $-138\pm12$ \\
S6 & $0.83\pm0.2$ & $0.46\pm0.2$ & $5.8\pm0.1$ & $7\pm4$ & $600\pm360$   & $0.67\pm0.2$ & $40\pm24$   & $-141\pm4  $ \\
\end{longtable}
\noindent

%============================================================
%========= Table 2
\renewcommand{\tabcolsep}{0.1cm}
\begin{longtable}{ccccccccccc}
\caption{Properties of Subregions  \label{tab:b}}\\\hline\hline

\centering
Region & $\delta V$ & $\delta \theta$ & $B_{\rm pos}$ (DCF) & $B_{\rm pos}$ (ST) & $\lambda_{\rm obs}$ (DCF)  & $\lambda_{\rm obs}$ (ST) & $\rm V_{A, 3D}^{DCF}$ & $\rm V_{A, 3D}^{ST}$ & $M_{\rm A}^{\rm DCF}$  & $M_{\rm A}^{\rm ST}$\\
           & ($\rm km~s^{-1}$) & (degree) &  ($\rm \mu G$) &  ($\rm \mu G$) &   &   & ($\rm km~s^{-1}$) & ($\rm km~s^{-1}$) & \\\hline 
\endfirsthead
\hline
\endfoot
\hline\\
\multicolumn{11}{c}{\parbox{\textwidth}{\textbf{Note. }Properties of the three subregions R1, R2, and R3. Columns 2 and 3 are the velocity dispersion and the dispersion of polarization angles. Columns 4--5 are the magnetic field strengths derived from the DCF and ST methods, respectively. Columns 6, 8 and 10 are the mass-to-flux ratio, 3-D $Alfv\acute{e}n$ velocity and Mach number calculated with the $B_{\rm pos}$ obtained from the DCF method. Columns 7, 9 and 11 denote the corresponding values computed based on the $B_{\rm pos}$ derived from the ST method}}
\endlastfoot
R1 & $0.64\pm0.4$  & $5.03$ & $101\pm71$ &  $42\pm27$  & $0.24\pm0.19$ &  $0.58\pm0.42$   & $4.62\pm3.3$  & $1.92\pm1.3$ &  $0.24\pm0.23$  &  $0.58\pm0.54$\\
R2 & $0.51\pm0.2$  & $7.57$ & $ 54\pm27$  &  $28\pm12$  & $0.45\pm0.27$ &  $0.87\pm0.46$   & $2.47\pm1.3$  & $1.28\pm0.6$ &  $0.36\pm0.24$  &  $0.69\pm0.42$\\
R3 & $0.40\pm0.2$  & $4.23$ &  $75\pm45$  &  $29\pm15$  & $0.32\pm0.22$ &  $0.84\pm0.51$   & $3.43\pm2.1$  & $1.33\pm0.7$ &  $0.20\pm0.16$  &  $0.52\pm0.38$\\
\end{longtable}
\noindent

\clearpage

%============================================================
%========= Appendix

\appendix 
\section{Distance of L914 Cloud} \label{sec:appendix1}
\begin{figure}
\centering
\includegraphics[scale=0.45]{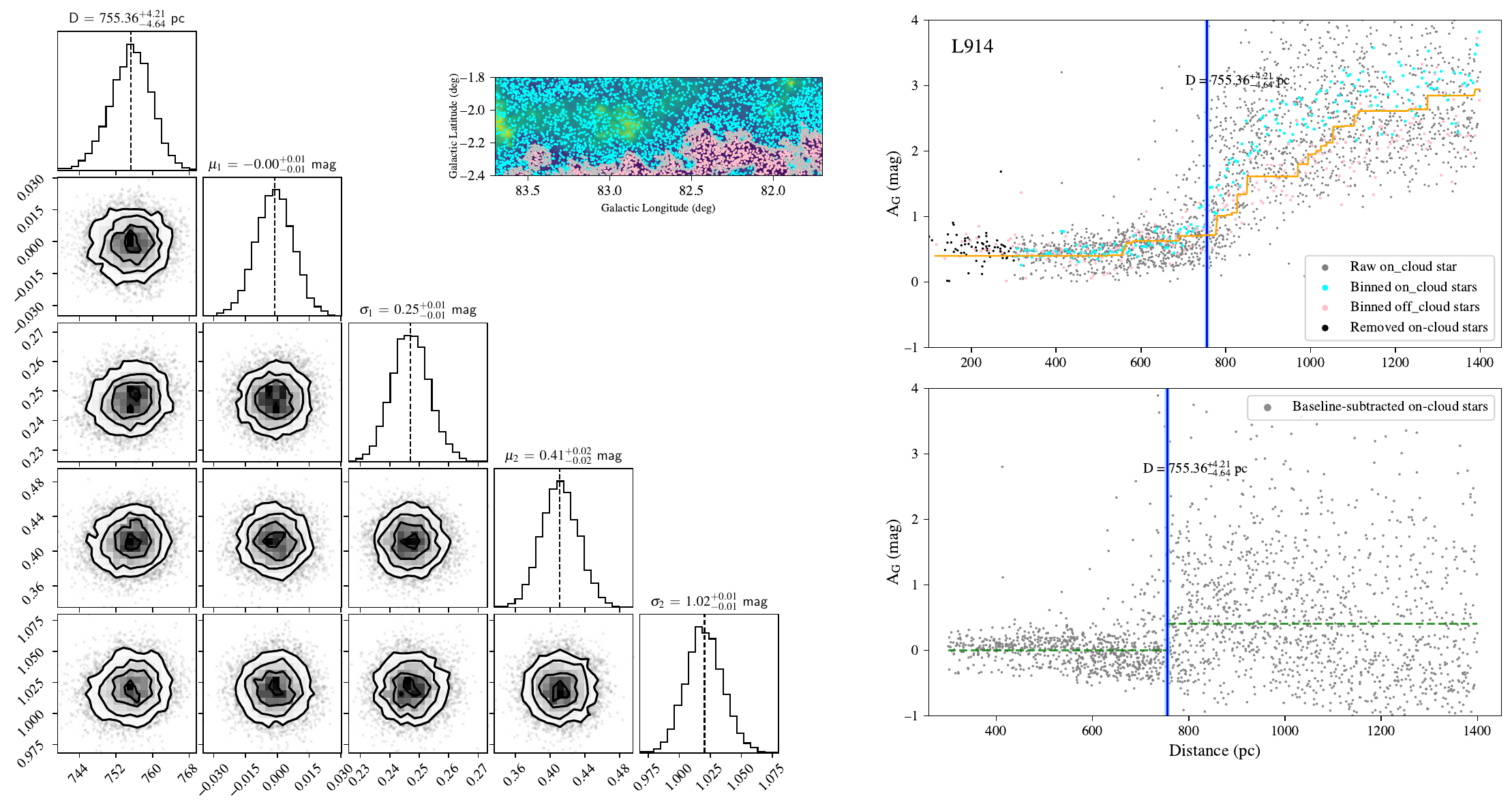}
\caption{Measured distance of L914. The middle panel shows the selected region used for distance estimation. The colorful background represents the MWISP $\rm ^{12}CO$ emission, and the grey contour shows the edge of the cloud ($\rm \sim 3\sigma$ ). The cyan and pink dots represent the on and off-cloud $Gaia$ DR3 stars, respectively. The corner plots of the Markov Chain Monte Carlo (MCMC) sampling (distance, the extinction values and uncertainties of foreground and background stars) are shown in the left panel, where the mean values of the samples are shown with solid vertical lines. In the right panels, the dots show the $Gaia$ stars as illustrated in the legend of each panel. The orange solid line in the upper right panel represents the result of monotonic regression fitting, while the dashed green lines in the bottom right panel are the modeled extinction $A_{\rm G}$.}
\label{app}
\end{figure}

In principle, because of the existence of molecular clouds along the line of sight, the extinction value $A_{\rm G}$ of on-cloud stars will produce a jump point, which can be used to deduce the distance of the cloud by using a Bayesian analysis. In this work, we adopt the same prior distribution and likelihood function as described in \citet{Yan2019} and then use the Python package emcee\footnote{\url{https://emcee.readthedocs.io/en/stable}} \citep{Daniel2013} to compute the posterior probability distribution of the cloud distance ($D$). The sampling result is shown in Figure~\ref{app} and the mean value of  $\rm \sim760~pc$ is adopted as the distance of the cloud. 

\end{document}